\documentclass[aps,noshowpacs,english,superscriptaddress,twocolumn]{revtex4-1}
\usepackage{amsmath,amssymb,amsfonts,amsthm,mathtools}
\usepackage{pgfplots,pgfplotstable,physics,epstopdf}
\usepackage{amsthm}
\usepackage{CJK}
\usepackage{bm}
\usepackage{babel}
\usepackage{graphicx}
\usepackage{subfigure}
\usetikzlibrary{shadings}
\usepackage[colorlinks=true,linkcolor=blue,anchorcolor=red,citecolor=blue,urlcolor=blue]{hyperref}
\usepackage{appendix}
\usepackage{tikz,pgf}
\usepackage{pdfpages}
\usepackage{verbatim}
\makeatletter
\AtBeginDocument{\let\LS@rot\@undefined}
\makeatother

\def \K {\hat{\mathcal{K}}}
\def \Z {\mathbb{Z}}
\def \H {\mathcal{H}}
\def \k {\bm{k}}

\def \T{\mathcal{T}}
\def \P{\mathcal{P}}
\def \C{\mathcal{C}}
\def \S{\mathcal{S}}
\def  \SS{\hat{\mathcal{S}}}
\def \TT {\hat{\mathcal{T}}}

\def \PP {\hat{\mathcal{P}}}

\def \Q {\mathcal{Q}}
\def \U {\mathcal{U}}
\def \O {\mathcal{O}}

\def \i {\mathrm{i}}

\def \I {\hat{I}}

\begin{document}

	\title{Takagi topological insulator with odd $\P\T$ pairs of corner states}
	
	\author{Jia-Xiao Dai}
	\email[These authors contributed  equally  to this work.]{}
	\affiliation{National Laboratory of Solid State Microstructures and Department of Physics, Nanjing University, Nanjing 210093, China}
	
	\author{Kai Wang}
	\email[These authors contributed  equally  to this work.]{}
	\affiliation{National Laboratory of Solid State Microstructures and Department of Physics, Nanjing University, Nanjing 210093, China}
	
	\author{Shengyuan A. Yang}
	\address{Research Laboratory for Quantum Materials, Singapore University of Technology and Design, Singapore 487372, Singapore}
	
	\author{Y. X. Zhao}
	\email[]{zhaoyx@nju.edu.cn}
	\affiliation{National Laboratory of Solid State Microstructures and Department of Physics, Nanjing University, Nanjing 210093, China}
	\affiliation{Collaborative Innovation Center of Advanced Microstructures, Nanjing University, Nanjing 210093, China}

\begin{abstract}
We present a novel class of topological insulators, termed the Takagi topological insulators (TTIs), which is protected by the sublattice symmetry and spacetime inversion ($\P\T$) symmetry. The required symmetries for the TTIs can be realized on any bipartite lattice where the inversion exchanges sublattices. The protecting symmetries lead to the classifying space of Hamiltonians being unitary symmetric matrices, and therefore Takagi's factorization can be performed. Particularly, the global Takagi's factorization can (cannot) be done on a $3$D ($2$D) sphere. In 3D, there is a $\mathbb{Z}_2$ topological invariant corresponding to the parity of the winding number of Takagi's unitary-matrix factor over the entire Brillouin zone, where the $\Z_2$ nature comes from the $O(N)$ gauge degrees of freedom in Takagi's factorization. In 2D, the obstruction for a global Takagi's factorization is characterized by another $\mathbb{Z}_2$ topological invariant, equivalent to the second Stiefel-Whitney number. For the third-order topological phases, the $3$D TTIs feature a parity condition for corner zero-modes, \emph{i.e.}, there always exist odd $\P\T$ pairs of corners with zero-modes. Moreover, for any $\P\T$ invariant sample geometry, all configurations of corner zero-modes satisfying the parity condition can exist with the same nontrivial bulk topological invariant. Actually, without closing the bulk gap, the boundary phase diagram have a cellular structure, where each topological boundary phase associated with a particular (cross-order) boundary-mode pattern corresponds to a contractible cell with certain dimension in the parameter space. 
\end{abstract}
\maketitle

\section{Introduction}
The discovery of symmetry-protected topological phases, such as topological insulators and superconductors, has attracted a broad interest during the last fifteen years~\cite{Kane-RMP,XLQi-RMP,ShinseiRyu-RMP,Volovik:book,Shun-Qing-Shen}. Theoretically, the topological band theory has been developed for characterizing the topological states, which was founded upon the topological $K$ theory~\cite{Atiyah-KR,Schnyder2008,Kitaev2009AIP,ZhaoYXWang13prl}. As a cornerstone, the  classification table for topological phases protected by time-reversal $\T$, particle-hole $\C$, and sublattice (chiral) $\S$ symmetries has been established by the real $K$ theory~\cite{AZ-Classification,Kitaev2009AIP,ZhaoYXWang13prl,ZhaoYXWang14Septprb}. There, for each symmetry class, the system is characterized by certain bulk topological invariant, and possesses topological boundary modes which are uniquely determined by the invariant and robust against any symmetry-preserving perturbations.
	
Two new directions emerged recently, which further broadens the scope of topological phases.
First, topological phases have been extended to ``higher order"~\cite{ZhangPRL2013,Benalcazar61,LangPRL2017,SongPRL2017}. For instance, for a three-dimensional ($3$D) system, we should explore boundary modes not only on the 2D surfaces (\emph{i.e.}, the first-order boundary) but also on the 1D hinges and 0D corners (\emph{i.e.}, the second- and the third-order boundaries, respectively). Second, spatial symmetries have been considered for enriching the topological phases. Particularly, for the spacetime-inversion symmetry $\P\T$, namely, the combination of $\T$ and the spatial inversion $\P$, a complete topological classification has been established by using the orthogonal $K$-theory~\cite{ZhaoWang16Aprprl}. The $\P\T$ symmetry is special in that it is a symmetry for every $k$-point of the reciprocal space, and hence it generates unique topological structures with many remarkable consequences,
such as nodal-loop linking structures, non-Abelian topological charges, and cross-order boundary phase transitions~\cite{Band-Combinatorics,ZhaoLu17Aprprl,PhysRevLett.115.036806,PhysRevLett.115.036807,PhysRevB.93.205132,B-J-Yang19APRPRX,ZhaoYang19prl,Wu1273,Wangzhijun2019prl,PhysRevLett.124.193901,AhnPRL2018}. 
	
In this paper, we reveal a new topological phase which is at the merging point of the two directions mentioned above. This phase is enabled by the $\P\T$ symmetry without spin-orbital coupling (SOC) and the sublattice symmetry  $\S$, with the anti-commutation relation
\begin{equation}
	\{\S,\P\T\}=0.
\end{equation}
This can be naturally realized on bipartite lattices with $\P$ exchanging sublattices. We show that for such systems, the Hamiltonians are restricted into the space of unitary symmetric matrices which permit Takagi's factorization, and topology is manifested in the resulting Takagi factor matrices.
In 3D, the $\Z_2$ topological invariant is the parity of the winding number of Takagi's factor over the Brillouin zone (BZ) and the $\Z_2$ nature is due to the $O(N)$ gauge degrees of freedom in Takagi's decomposition~\cite{Takagi,Takaji_wiki}. In 2D, the $\Z_2$ invariant characterizes the obstruction for a global factorization.

This new class of topological insulators are termed as Takagi topological insulators (TTIs). Remarkably, we find that the $3$D TTIs satisfy a parity condition: its third-order phases always have \emph{odd} $\P\T$ pairs of corners with zero-modes. Moreover, with the same nontrivial bulk topological invariant, for any given geometry, all possible configurations of corner zero-modes can exist, as long as they satisfy the parity condition. Without closing the bulk gap, high-order topological phases for a TTI are intermediated by lower-order topological phases.
Specifically, the boundary phase diagram features a cellular structure  for symmetry-preserving perturbations: Each open cell in the parameter space corresponds to a certain boundary-mode distribution with a unique or mixed boundary order.
For instance, for a cubic shape,  there are two distinct patterns of corner states, \emph{i.e.}, two zero-modes located at a single pair of antipodal corners or six zero-modes at three such pairs. A phase transition between them occurs with zero-modes on mixed-order boundaries, involving both corner modes and helical hinge modes.

\section{Symmetries and topological invariants}
   We start with the elementary symmetries to be considered. For spinless systems, the $\T$ operator in momentum space is $\TT=\K\I$, with $\K$ the complex conjugation and $\I$ the inversion of momenta. Assuming a bipartite lattice where $\P$ inverses the two sublattices, namely $\PP=\sigma_1\I$~\cite{PT-symmetry}, then the sublattice symmetry $\S$ is represented by $\SS=\sigma_3$. Here, $\sigma_i$ with $i=1,2,3$ are the Pauli matrices. 
   Hence, the inversion naturally anti-commutes with $\S$: $\{\PP,\SS\}=0$, so does $\PP\TT$, with $\{\PP\TT,\SS\}=0$. In this paper, in addition to $\S$, we only assume the combined symmetry $\P\T$, while individual $\P$ and $\T$ can both be violated.

   These two symmetries put constraints on the form of the Hamiltonian $\H(\k)$ in momentum space. $\S$ requires $\H(\k)$ to be block anti-diagonal, \emph{i.e.},
   \begin{equation}\label{anti-diagonal}
   	\widetilde{\H}(\k)=\begin{bmatrix}
   		0 & \Q(\k)  \\
   		\Q^\dagger(\k) & 0
   	\end{bmatrix},\quad ~\Q\Q^\dagger=I_N,
   \end{equation}
Here, for topological study of gapped phases, we have flattened the Hamiltonian to be $\widetilde{\H}=\mathrm{sgn}(\H)$, \emph{i.e.}, $\H(\k)$ is adiabatically deformed into $\widetilde{\H}(\k)$ with all conduction/valence bands having energy $\pm 1$, and $N$ is the number of valence (conduction) bands. It follows that $\Q$ is unitary.
Substituting $\hat{\P}\hat{\T}=\sigma_1\K$ into the symmetry condition, $[\hat{\P}\hat{\T},\widetilde{\H}(\k)]=0$, we can find that $\Q(\k)$ is symmetric.  
Thus, $\Q(\k)$ is constrained to be a unitary symmetric matrix for each $\k$, \emph{i.e.},
\begin{equation}
	\Q\Q^\dagger=I_N,\quad \Q^T(\k)=\Q(\k).
\end{equation}
   As shown in Appendix. \ref{A}, for symmetric unitary matrices, Takagi's factorization can be performed on $\Q(\k)$~\cite{Takagi,Takaji_wiki}:
   \begin{equation}\label{Takagi}
   \Q(\k)=\U(\k)\U^T(\k),
   \end{equation}
   where $\mathcal{U}(\k)\in U(N)$ is a unitary matrix. It is crucial to note that there is a gauge degree of freedom in the factorization, \emph{i.e.}, $\Q(\k)$ is invariant under the gauge transformation:
   \begin{equation}\label{Gauge-Transf}
   \U(\k)\mapsto \U(\k) \O(\k),
   \end{equation}
   where $\O(\k)\in O(N)$ is an orthogonal matrix satisfying $\O(\k)\O^T(\k)=I_N$. Therefore, the classifying space for this symmetry class is
   \begin{equation}
   US(N)=U(N)/O(N).
   \end{equation}

   Since $\pi_3[US(N)]\cong \Z_2$, the class allows a $\Z_2$ topological classification in 3D. We proceed to analyze the topological structure and formulate the corresponding $3$D $\Z_2$ invariant. For simplicity, let's first assume that the base space is a $3$D sphere $S^3$. It is tempting to lift the distribution $\Q(\hat{d}_\mu)$ of symmetric unitary matrices over $S^3$ to a distribution $\U(\hat{d}_\mu)$  of unitary matrices, where $\hat{d}_\mu$ denotes a point on $S^3$. However, it is not yet clear whether such a lifting exists globally. Recall that for Chern insulators in class A, the flattened Hamiltonian can be expressed as $\widetilde{\mathcal{H}}(\k)=\mathcal{T}(\k)\Lambda \mathcal{T}^\dagger(\k)$, with $\Lambda=\mathrm{diag}(I_M,-I_N)$ and $\mathcal{T}(\k)\in U(M+N)$ ($M$ and $N$ are the numbers of conduction and valence bands, respectively). The global lifting of $\widetilde{\mathcal{H}}$ into $\mathcal{T}$ over the whole $2$D BZ is impossible for Chern insulators, and the Chern number actually characterizes this obstruction~\cite{bernevig2013topological}.

   \begin{figure}
   	\includegraphics[scale=0.4]{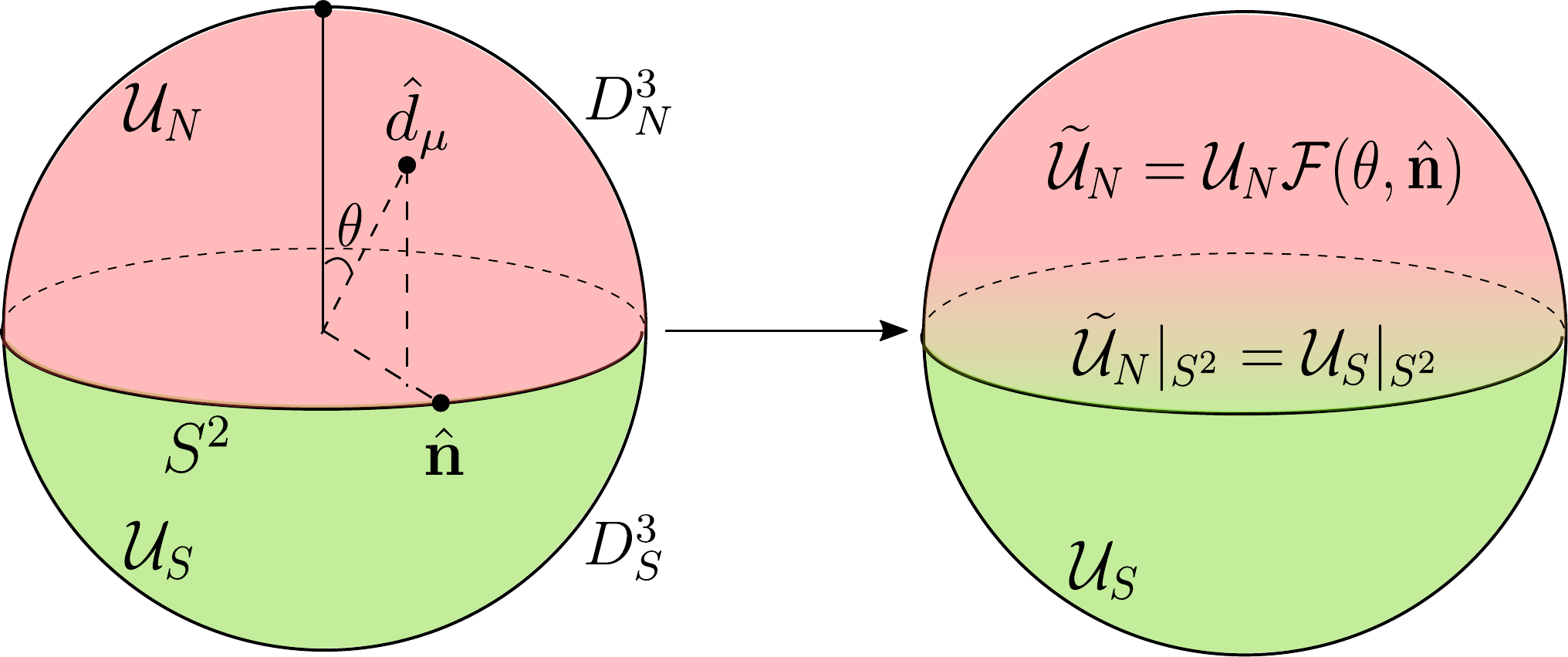}
   	\caption{\label{fig:Lifting} Existence of global lifting. The liftings $\U_{N,S}$ over the north and south hemispheres $D_{N,S}^3$ are smoothened into a global lifting by modifying $\U_{N}$ as $\widetilde{\U}_{N}=\U_{N}\mathcal{F}(\theta,\hat{\bm{n}})$. }
   \end{figure}

   In the present case, we show that such a lifting always exists. We cover $S^3$ by the north and south hemispheres, $D^3_N$ and $D^3_S$, with the intersection being the equator $S^2=D^3_N\cap D^3_S$ (see Fig.~\ref{fig:Lifting}). Since $D^3_{N,S}$ are contractible, the corresponding liftings $\U_{N,S}$ over them are always possible. Restricted to each point $\hat{\bm{n}}$ on the equator $S^2$, $\U_{N}$ and $\U_{S}$ correspond to the same symmetric unitary matrices $\Q|_{S^2}$, and hence, they must be related to each other through $\U_{N}(\hat{\bm{n}})\O_{S^2}(\hat{\bm{n}})=\U_{S}(\hat{\bm{n}})$ with $\O_{S^2}(\hat{\bm{n}})\in O(N)$ for any point $\hat{\bm{n}}$ on the equator $S^2$.  Then, since $\pi_2[O(N)]=0$, one can always smoothly deform $\O_{S^2}$ into a constant function. In other words, there exists a smooth two-variable function $\mathcal{F}(\theta,\hat{\bm{n}})\in O(N)$ with $\theta\in [0,\pi/2]$, such that $\mathcal{F}(\pi/2,\hat{\bm{n}})=\O_{S^2}(\hat{\bm{n}})$ and $\mathcal{F}(0,\bm{n})$ is constant. Interpreting $\theta$ as the inclination (polar angle) of $\hat{d}_\mu\in S^3$ (see Fig.~\ref{fig:Lifting}), namely $\hat{d}_\mu=(\theta,\hat{\bm{n}})$, we can modify the lifting $\U_N$ as $\widetilde{\U}_N(\hat{d}_\mu)=\U_N(\hat{d}_\mu)\mathcal{F}(\theta,\hat{\bm{n}})$. $\widetilde{\U}_N(\hat{d}_\mu)$ gives the same $\Q|_{D^3_N}$ according to \eqref{Takagi}, but now $\widetilde{\U}_N$ is continuously connected to $\U_{S}$ on the equator.  Thus, we can always smoothen $\U_{N,S}$ to obtain a global lifting $\U$ over $S^3$.


   Based on the above discussion, we can assume a global lifting $\U(\k)$ over the whole 3D BZ~\cite{Global-Torus_Lifting}. Then, we claim that the 3D $\Z_2$ invariant for our symmetry class can be written as
   \begin{equation}
   \nu=\frac{1}{24\pi^2}\int_\text{BZ} d^3k~\epsilon^{ijk}\text{tr}\ \U\partial_{i} \U^\dagger \U\partial_{j} \U^\dagger \U\partial_{k} \U^\dagger~ \mathrm{mod~}2,\label{invariant}
   \end{equation}
   where $\epsilon^{ijk}$ is the antisymmetric tensor. Without the modulo 2 operation, the expression gives a $\Z$ invariant because $\pi_3[U(N)]\cong \Z$~\cite{Comparison_AIII}.
   Here, the invariant is reduced from $\Z$ to $\Z_2$. This is because a topological invariant for $\Q$ should be unchanged under any gauge transformation~\eqref{Gauge-Transf}.
   Straightforward derivations show that
   \begin{equation}
   \nu[\U\O]=\nu[\U]+\nu[\O].
   \end{equation}
   Substituting a topologically nontrivial $\O$ into the formula \eqref{invariant} gives an even integer, namely $\nu[\O]\in 2\Z$. Thus, only the parity of the integer is meaningful here. This justifies the $\Z_2$ nature of the topological invariant \eqref{invariant}.

   The gauge freedom in Takagi's factorization [Eq.~\eqref{Gauge-Transf}] also enables a $2$D $\Z_2$ invariant. Consider a $2$D sphere $S^2$, which is divided into north and south hemispheres $D^2_{N,S}$ overlapping along the equator $S^1$. Again, because of the gauge freedom, the Takagi factors $\U_{N,S}$ over $D^2_{N,S}$, respectively, are equal on the equator only up to a gauge transformation $\O_{S^1}$. However, the fundamental group of $O(N)$ is nontrivial: $\pi_1[O(N)]\cong \Z_2$ for $N>2$, which leads to obstructions for a global Takagi's factorization over $S^2$ and meanwhile gives a $\Z_2$ classification for 2D.

   The $\Z_2$ classification revealed here defines TTIs in 3D and 2D. It should be mentioned that the origins of the invariants for 2D and 3D are quite different. For 3D, the invariant corresponding to $\pi_3[US(N)]$ for the globally lifted Takagi factor, whereas for 2D, the invariant corresponds to $\pi_1[O(N)]$ characterizing the obstruction to the global lifting. The 2D case resembles the origin of the monopole charge, such as the first Chern number and the second Stiefel-Whitney number for $\P\T$-invariant systems~\cite{ZhaoLu17Aprprl,Stiefel-Whitney}. In the Appendix. \ref{B}, we show that the $2$D invariant is exactly equivalent to the second Stiefel-Whitney number that does not require the sublattice symmetry.
   Hence, even with the sublattice symmetry violated, nontrivial topology of a 2D TTI can still be present as long as $\P\T$ is preserved. In contrast, for $3$D TTIs, the sublattice symmetry is essential for the nontrivial topology.

   \section{Model for 3D TTI}  We construct a concrete Dirac model for a 3D TTI.  To preserve both $\S$ and $\P\T$, we need four Dirac matrices commuting with $\PP\TT=\sigma_1\K$ and anti-commuting with $\SS=\sigma_3$, of which three are for the kinetic terms and one is for the mass term. The minimal dimension required is eight. Let $\Gamma^A$ with $A=1,2,\cdots,7$ be the $8\times 8$ Dirac matrices satisfying $\{\Gamma^A,\Gamma^B\}=2\delta^{AB} I_8$. We choose the explicit representation:
   $\Gamma^\mu=\sigma_1\otimes\gamma^\mu$, with $\mu=1,\cdots, 4$, $\Gamma^5=\sigma_1\otimes\gamma^5$, $\Gamma^6=\sigma_3\otimes I_4 $, and $\Gamma^7=\sigma_2\otimes I_4$. Here, $\gamma$'s are the $4\times 4$ Dirac matrices:
   $\gamma^1=\tau_1\otimes\kappa_1,\gamma^2=\tau_1\otimes\kappa_3,\gamma^3=\tau_3\otimes\kappa_0,\gamma^4=\tau_1\otimes\kappa_2$, $\gamma^5=\tau_2\otimes\kappa_0$. $\sigma_i$, $\tau_i$ and $\kappa_i$ are three sets of Pauli matrices. Hereafter, the Roman subscript $i$ runs from 1 to 3, whereas the Greek subscripts run from 1 to 4.

   It is straightforward to check that we can use $\Gamma^i$ with $i=1,2,3$ for kinetic terms and $\Gamma^7$ for the mass term. Consequently, the Hamiltonian is
   \begin{equation}\label{Hamiltonian}
   \H_0=\bm{d}(\k)\cdot \bm{\Gamma}+d_4(\k) \Gamma^7,
   \end{equation}
   where $d_i(\k)=t\sin k_i$ and $d_4(\k)=M-\lambda \sum_{i=1}^3\cos k_i$. The spectrum is given by $E(\k)=\pm |d_\mu(\k)|$ with $ |d_\mu|=(\sum_\mu d_\mu^2)^{1/2}$.  For insulating states with $|d_\mu(\k)|>0$, the flattened Hamiltonian is $\widetilde{\H}_0=\hat{\bm{d}}\cdot\bm{\Gamma}+\hat{d}_4\Gamma^7$, with $\hat{d}_\mu=d_\mu/|d_\mu|$. In such a case,
   \begin{equation}\label{Dirac_Q}
   \Q(\hat{d}_\mu)=-\i \exp (\i \hat{\bm{n}}\cdot \bm{\gamma}~\theta ),
   \end{equation}
   where $\cos \theta=\hat{d}_4$ and $\sin \theta =|\bm{d}|/|d_\mu|$, namely, $\theta$ is the inclination of $\hat{d}_\mu \in S^3$. Since $\gamma^i$'s are symmetric, $\Q$ is indeed a symmetric unitary matrix.

   To obtain a global Takagi's factorization \eqref{Takagi}, we introduce
   \begin{equation}
   \U_N(d_\mu)=e^{\i \frac{3\pi}{4}} e^{\i\hat{\bm{n}}\cdot \bm{\gamma}\frac{\theta}{2}},\quad  \U_S(d_\mu)=e^{\i \frac{5\pi}{4}} e^{\i  \hat{\bm{n}}\cdot \bm{\gamma}\frac{\theta+\pi}{2}}.
   \end{equation}
   Here, $\U_N$ ($\U_S$) is the factorization for $\Q$ \eqref{Dirac_Q} on the north (south) hemisphere with $\theta\in [0,{\pi}/{2}]$ ($[{\pi}/{2},\pi]$). On the equator $S^2$ with $\theta=\pi/2$, they are related by $\U_N(\hat{\bm{n}})\O(\hat{\bm{n}})=\U_S(\hat{\bm{n}}) $ with $\O(\hat{\bm{n}})=-\hat{\bm{n}}\cdot\bm{\gamma}$.
   A smooth deformation of $\O(\hat{\bm{n}})$ to a constant function can be chosen as
   \begin{equation}
   	\mathcal{F}(\theta,\hat{\bm{n}})=\cos\theta \gamma^4\gamma^5-\sin \theta \hat{\bm{n}}\cdot\bm{\gamma},
   \end{equation}
   with $\theta\in [0,\pi/2]$. It is straightforward to check that $\mathcal{F}(\theta,\hat{\bm{n}})$ is a path in $O(N)$ for each $\hat{\bm{n}}\in S^2$. Thus, a global Takagi's factorization $\U(d_\mu)$ is given by $\widetilde{\U}_N(d_\mu)=\mathcal{U}_N(\theta,\hat{\bm{n}})\mathcal{F}(\theta,\hat{\bm{n}})$ on the north hemisphere and $\mathcal{U}_S(\theta,\hat{\bm{n}})$ on the south hemisphere. Substituting this global $\U$ into Eq.~\eqref{invariant}, we obtain
   \begin{equation}
   \nu=\frac{1}{2\pi^2}\int_\text{BZ} d^3k~\epsilon^{\mu\nu\lambda\sigma} \hat{d}_\mu \partial_1 \hat{d}_\nu \partial_2 \hat{d}_\lambda \partial_3 \hat{d}_\sigma \mod 2.
   \end{equation}
   If we set $t=\lambda=1$, the model \eqref{Hamiltonian} is in the TTI phase if $1<|M|<3$. 

\begin{figure}
	\includegraphics[scale=0.4]{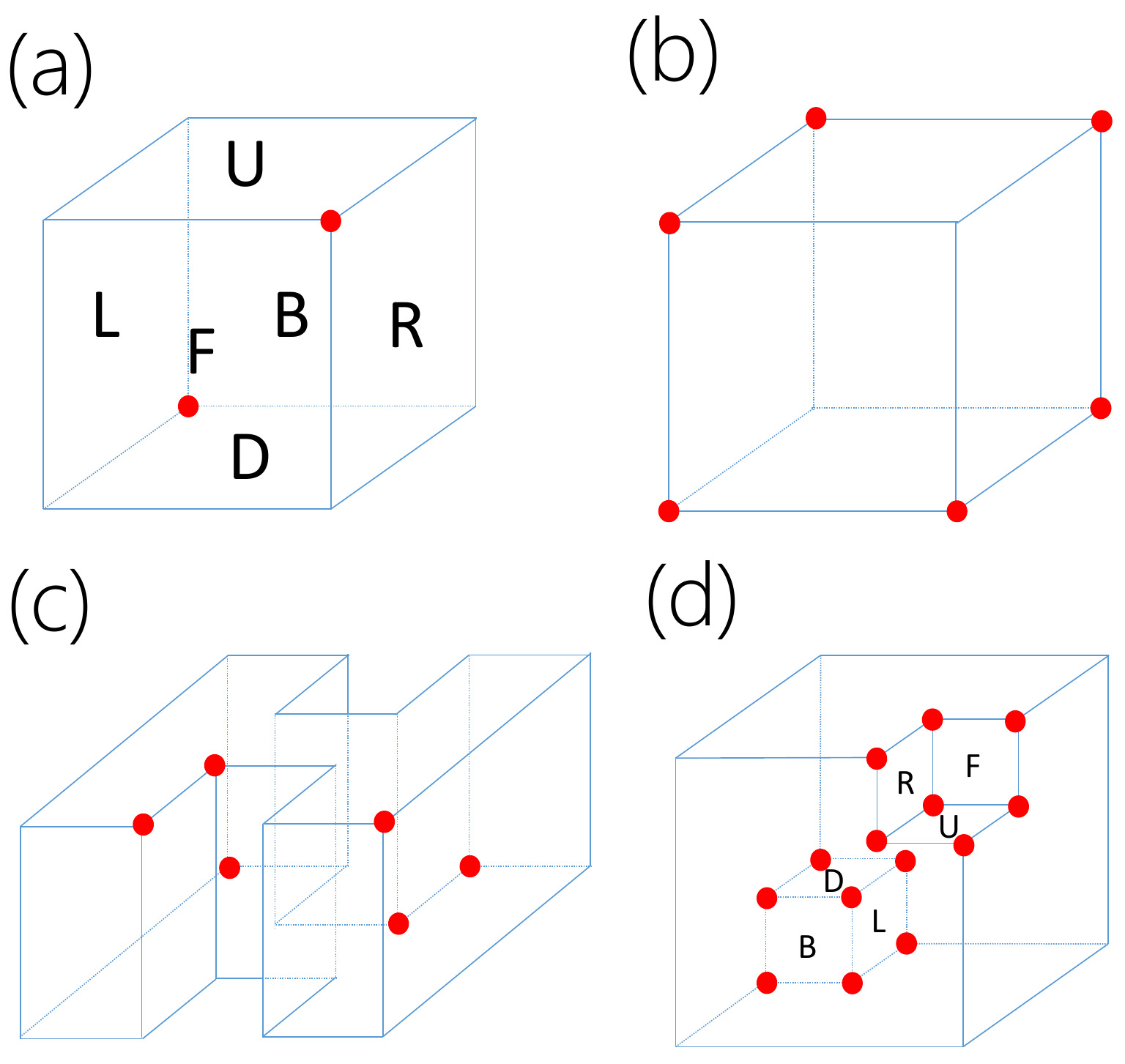}
	\caption{Configurations of corners with zero-modes. Corners with zero-modes are marked by red dots. The number of $\P\T$ pairs is always odd.}
	\label{Oddpairs}
\end{figure}
	
  \section{Odd $\P\T$ pairs of corners with zero-modes}
The model $\H_0(\k)$ in \eqref{Hamiltonian} describes a first-order topological insulator with $2$D Dirac surface modes on all surfaces. But there are other symmetry-preserving terms. We group those that can affect the surface states as
  \begin{equation}\label{pert}
  	\Delta\H_1=\sum_{i=1}^3\sum_{\alpha=4,5}\i\lambda_{i\alpha}\Gamma^i\Gamma^\alpha \Gamma^7,~~ \Delta\H_2=\sum_{i=1}^3\i\eta_{i}\Gamma^i\Gamma^6.
  \end{equation}
 As shown in Appendix. \ref{C}, all other symmetry-preserving perturbations are irrelevant to the boundary modes. The terms in (\ref{pert}) span a 9-d parameter space, in which the boundary modes are studied by both analytical and numerical approaches. Below, we address the key features of the phase diagram, while relegating the calculation details to the Appendix. \ref{D}.

  The most generic phases of a $3$D TTI are third-order topological phases, which are stable under all symmetry-preserving perturbations. For a given $\P\T$-invariant geometry, the nontrivial topological invariant corresponds to a variety of configurations regarding zero-mode distribution [see Fig.~\ref{Oddpairs}(a) and (b) for a cubic geometry]. Importantly, there is a parity condition, \emph{i.e.}, a configuration can be realized if and only if the number of $\P\T$ pairs is odd, as illustrated in Fig.~\ref{Oddpairs} and demonstrated in detail in Appendix. \ref{E}. In other words, a TTI features odd $\P\T$ pairs of corners with zero-modes, which is distinct from all previously known third-order topological insulator states. A detailed derivation can be found in the Appendix. \ref{E}.
  
\begin{figure}
  	\centering
  	\includegraphics[width=3.4in]{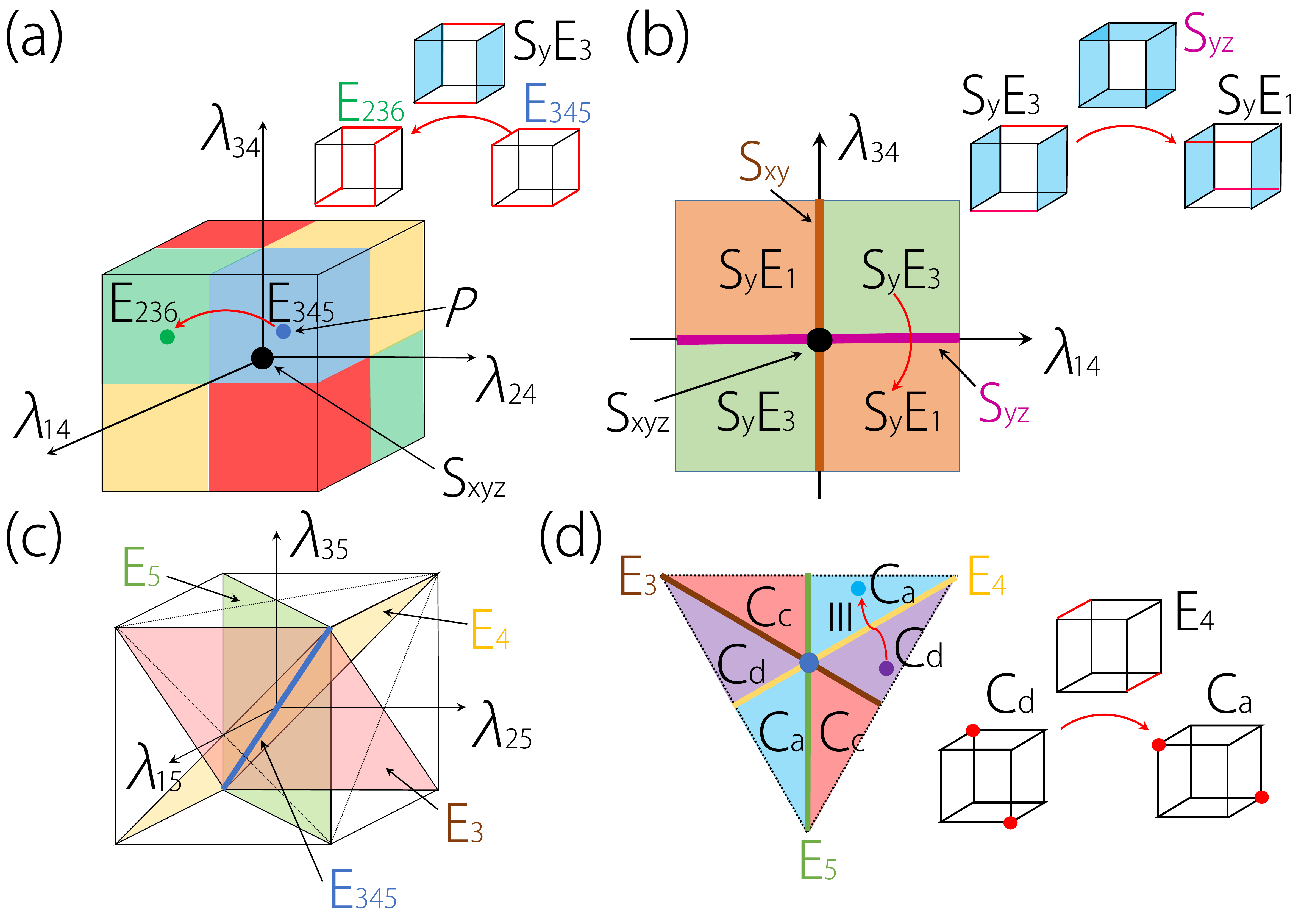}
  	\caption{The cellular structure of the boundary phase diagram under perturbation $\Delta\H_1$. Each cell is denoted by $A_{\xi}$, where $A=S, E, C$ stand for surface, edge and corner states, respectively, and $\xi$ labels the corresponding zero-mode distribution pattern. The naming scheme and a complete list can be found in the Appendices. \ref{F} and \ref{G}.  (a) Cellular phase diagram for $\lambda_{i4}$. (b) Cellular structure enriched by $\lambda_{i5}$ for a second-order TI phase. \label{phasediagram-Fig}
  	}
  \end{figure}

\begin{figure}
  	\centering
  	\includegraphics[width=3.3in]{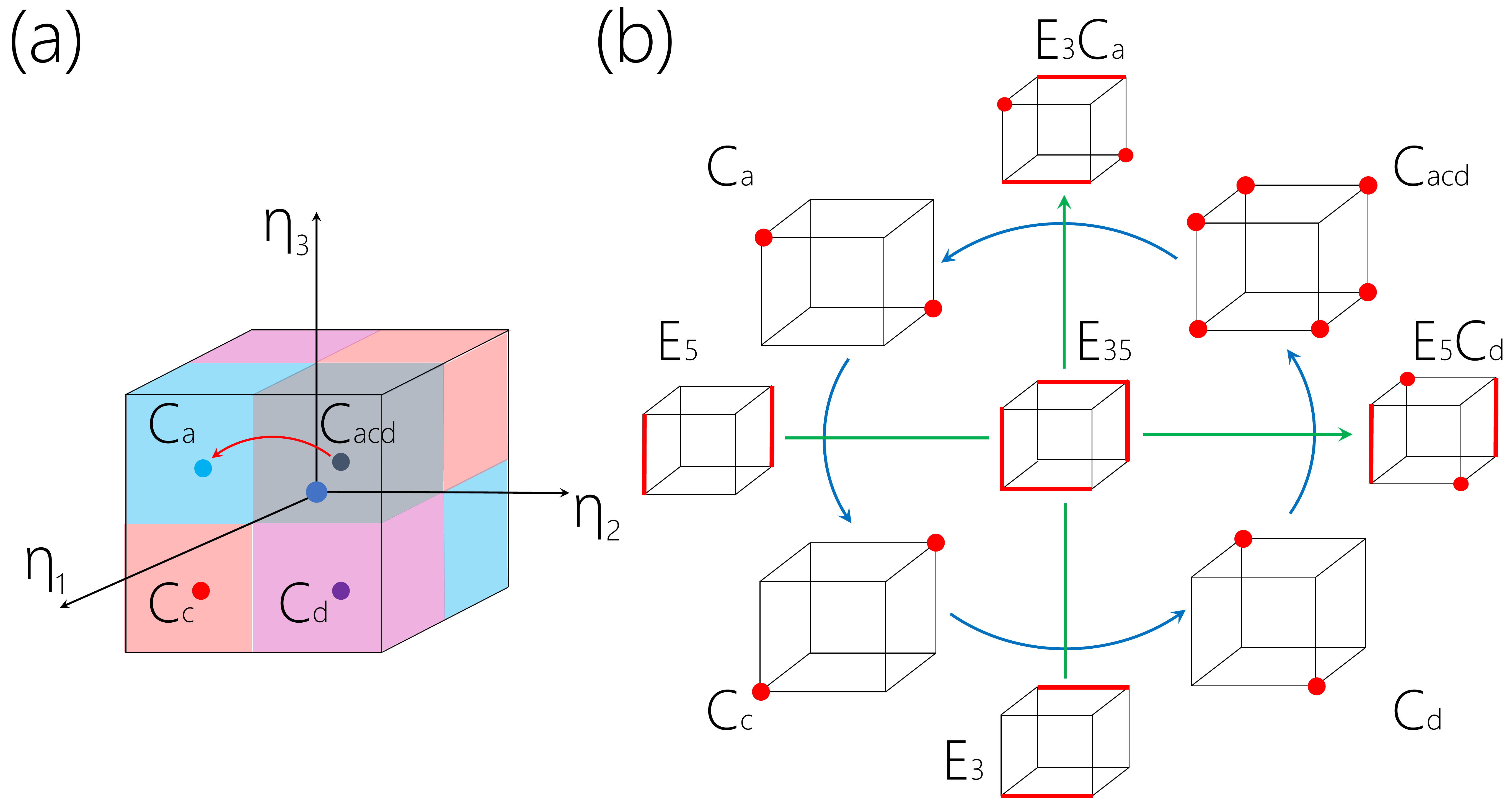}
  	\caption{The cellular structure enriched by $\Delta\H_2$ from a second-order TI phase. \label{phasediagram-2}
  	}
  \end{figure}

  Furthermore, without closing the bulk gap, these third-order topological phases are connected by intermediate cross-order boundary states (see Figs.~\ref{phasediagram-Fig} and \ref{phasediagram-2}), and the boundary phase diagram features a cellular structure. Below, we address the cellular boundary phase diagram and phase transitions on a cube-shaped sample.

  Let's first consider the three terms with parameters $\lambda_{i4}$ ($i=1,2,3$). The corresponding  phase diagram is shown in Fig.~\ref{phasediagram-Fig}(a). Obviously, the $0$-d cell at the origin corresponds to the first-order insulator phase of $\mathcal{H}_0$. The eight octants are $3$-d cells, of which each $\P\T$-related pair corresponds to a second-order topological-insulator phase with six gapless edges [Fig.~\ref{phasediagram-Fig}(a)]. Note that there are exactly four possible patterns of six connected edges that preserve the $\P\T$ symmetry. The eight $3$-d cells are attached on the $2$-d skeleton which is the union of the $x$-$y$, $y$-$z$ and $z$-$x$ planes. Excluding the $0$-d cell, each plane consists of four $2$-d cells, namely, the four quadrants, attached on the four $1$-d cells. For example, on the $x$-$z$ plane,  each $2$-d cell corresponds to a pattern of mixed-order boundary modes, with two $x$-$z$ gapless surfaces connected by two $\P\T$-related gapless $y$-edges [Fig.~\ref{phasediagram-Fig}(b)], while each $1$-d cell corresponds to gapless surfaces parallel to the $x$ or $z$ direction [Fig.~\ref{phasediagram-Fig}(b)]. The direct phase transition from a $n$-d cell to a neighboring $n$-d cell must go through a $(n-1)$-d cell in the phase diagram.
 For instance, starting with the first quadrant cell with two gapless off-diagonal $x$-edge connecting $y$-$z$ surfaces in Fig.~\ref{phasediagram-Fig}, the intermediate $1$D cell corresponds to the gapped $y$-$z$ surfaces and the gapless $x$-$z$ surfaces, which can be continued to gapless diagonal $x$-edges connecting $y$-$z$ surfaces.

  The cellular structure is further enriched when adding the three terms with $\lambda_{i5}$ in $\Delta \H_1$. For instance, if we focus on a point $P$ in the first octant in Fig.~\ref{phasediagram-Fig}(a), the additional three terms give rise to a cellular phase diagram as illustrated in Fig.~\ref{phasediagram-Fig}(c). The point $P$ extends into a $1$-d cell, namely, the diagonal line. Six $2$-d cells are attached onto the $1$-d cell, corresponding to the three shaded diagonal planes in Fig.~\ref{phasediagram-Fig}(c). Each $2$-d cell features a pair of $\P\T$-related gapless edges [Fig.~\ref{phasediagram-Fig}(d)]. Then, six $3$-d cells are attached onto this $2$-d skeleton, and each corresponds to a third-order phase with only two zero-modes at a pair of antipodal corners [Fig.~\ref{phasediagram-Fig}(d)].
  The critical point between two third-order phases is a second-order phase with a pair of $\P\T$-related gapless edges [Fig.~\ref{phasediagram-Fig}(d)].

  Similar analysis can be performed for $\Delta \H_2$ terms. Again, consider point $P$ in Fig.~\ref{phasediagram-Fig}(a). It grows into a cellular structure under the $\Delta \H_2$ terms [Fig.~\ref{phasediagram-2}(a)]. The boundary phases corresponding to the cells and the phase transitions between them are illustrated in Fig.~\ref{phasediagram-2}.
  Note that the most stable phases, the third-order topological insulators with one pair or three pairs of corner zero-modes, both correspond to 9-d cells in the complete parameter space. The direct transition between them corresponds to a mix-order phase with the pair of antipodal zero-mode corners and a pair of $\P\T$-related helical edges.

\section{Discussion}
Besides materials with negligible SOC on a bipartite lattice, TTIs could be readily realized in artificial systems, such as photonic/phononic crystals, electric circuit networks and mechanical arrays~\cite{Lu2014,YangZjprl2015,Mittal2019,Xue2020,Ronny_2018np,Yu_Zhao_NSR,Huber2020np}.  The symmetry relation $(\P\T)^2=1$ is inherent in these systems. 
The $2$D and $3$D topological invariants formulated here can also be applied to study $\P$-invariant topological superconductors in class CI (such as $d_{x^2-y^2}$ or $d_{xy}$ wave superconductors)~\cite{Schnyder2008}. Now the chiral symmetry is $\i\C\T$, rather than $\S$.

\begin{acknowledgements}
	This work is supported by National Natural Science Foundation of China (Grants No.
	11874201 and 12174181) and the Singapore Ministry of Education AcRF Tier 2
	(MOE2019-T2-1-001).
\end{acknowledgements}
\appendix

\section{Takagi's factorization}\label{A}
For a square, complex and symmetric matrix $A$ satisfying that $A=A^T$, there must exist a unitary matrix $V$ and a nonnegative diagonal matrix $D$ such that
\begin{equation}
	A=VDV^T,\label{Takagii}
\end{equation}
where $V^T$ is the matrix transpose of $V$. The diagonal elements of $D$ are the nonnegative square roots of the eigenvalues of $AA^\dagger$.\\

The decomposition (\ref{Takagii}) can be interpreted in two different ways.
First, we consider the eigenvalue decomposition of Hermitian matrices. The matrix $C=A^\dagger A$ is Hermitian and positive semi-definite, so there exists a unitary matrix $U$ such that $U^\dagger C U$ is diagonal with non-negative real diagonal elements. Thus, $Z=U^TAU$ satisfies that
	\begin{equation}
		Z^T=U^TA^TU=Z,~Z^\dagger Z=U^\dagger CU.
	\end{equation}
	In other words, $Z$ is complex symmetric and $Z^\dagger Z$ is real. Decomposing $Z=X+iY$ with $X$ and $Y$ real and symmetric, we can get that
	\begin{equation}
		Z^\dagger Z=X^2+Y^2+i[X,Y].
	\end{equation}
	Thus, $X$ commutes with $Y$ and there is a real orthogonal matrix $W$ such that both $WXW^T$ and $WYW^T$ are diagonal. Setting $V_1=UW^T$ ($V_1$ is unitary), the matrix $V_1^TAV_1$ is complex diagonal, which can be written as
	\begin{equation}
		V_1^TAV_1=\text{diag}(r_1e^{i\theta_1},r_2e^{i\theta_2},\cdots,r_ne^{i\theta_n}),
	\end{equation}
	with $r_i>0$ is real. Giving another diagonal matrix as $N=\text{diag}(e^{-i\theta_1/2},e^{-i\theta_2/2},\cdots,e^{-i\theta_n/2})$, $V^*=V_1N$ is also unitary and satisfies that
	\begin{equation}
		D=V^\dagger AV^*=\text{diag}(r_1,r_2,\cdots,r_n)~\Longrightarrow A=VDV^T.
	\end{equation}
	Since $D^2=D^\dagger D=(V^TA^\dagger V)(V^\dagger A V^*)=V^T A^\dagger AV^*$, $r_i^2$'s correspond to the eigenvalues of $C$. Thus, the decomposition (\ref{Takagii}) is proved.
	
	Next, we consider the singular value decomposition. For any $m\times n$ complex matrix $B$, there exists a singular value decomposition as $$B=VDU^\dagger.$$ $V$ is an $m\times m$ unitary matrix, $U$ is an $n\times n$ unitary matrix and $D$ is an $m\times n$ rectangular diagonal matrix with non-negative real numbers on the diagonal. Also, we know that
	\begin{equation}
		\begin{split}
			BB^\dagger=(VDU^\dagger)(UD^\dagger V^\dagger)=V(DD^\dagger )V^\dagger,\\
			B^\dagger B=(UD^\dagger V^\dagger)(VDU^\dagger)=U(D^\dagger D )U^\dagger.
		\end{split}
	\end{equation}
	Therefore, the diagonal elements of $D$ are the square roots of the non-zero eigenvalues of $BB^\dagger$ or $B^\dagger B$. If $B$ is square and symmetric, i.e., $m=n$ and $B=B^T$, we can know that
	\begin{equation}
		V^T=U^\dagger ~\Longrightarrow ~B=VDV^T.
	\end{equation}
	The decomposition (\ref{Takagii}) is proved again.
If $A$ is unitary, $D$ is the identity matrix. In other words, $A$ can always be decomposed into $A=VV^T$ if $A$ is a symmetric unitary matrix.
\section{The equivalence of 2D topological invariants}\label{B}
\begin{figure}[t]
	\includegraphics[width=1.8in]{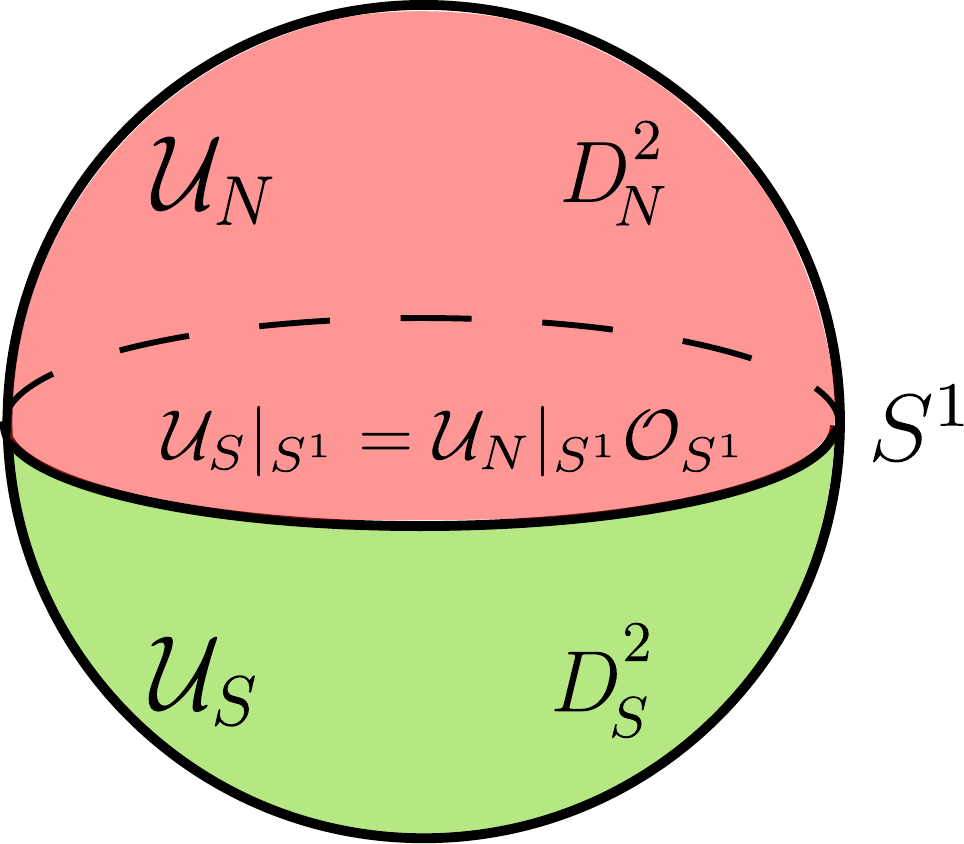}
	\caption{The Takagi factors in $2$D.}
	\label{CI2}
\end{figure}

As the main text explains, the flattened Hamiltonian $\widetilde\H(\k)$ with $\PP \TT$ and $\S$ symmetries can be represented as
\begin{equation}
	\widetilde{\H}(\k)=\begin{bmatrix}
		0&\Q(\k)\\
		\Q^\dagger(\k)&0
	\end{bmatrix},
	\quad \Q(\k)=\U(\k)\U^T(\k).\label{Takagi_2D}
\end{equation}
The symmetry operators are given as $\P \T=\sigma_1\K$ and $\SS=\sigma_3$. Consider a $2$D sphere $S^2$, which is divided into the north and south hemispheres $D^2_{N,S}$, overlapping along the equator $S^1$. The Takagi factors $\U_{N/S}$ over $D_{N/S}^2$, respectively, can be transformed to each other by a gauge transformation $\mathcal O_{S^1}$ over the intersection $S^1$ of two hemispheres, as shown in Fig.~\ref{CI2}. 
$\mathcal O_{S^1}$ is given by $$\mathcal O_{S^1}=\U^\dagger_N|_{S^1}\U_S|_{S^1}, ~\mathcal O_{S^1}\in \mathrm O(M).$$ $\pi_1[\mathrm{O}(M)]=\Z_2$ for $M>2$ leads to obstructions for a global Takagi's factorization over $S^2$.

Assuming the matrix dimension of $\widetilde\H(\k)$ in Eq. (\ref{Takagi_2D}) is $2M$, valence wavefunctions of $\widetilde\H(\k)$ can be given by
\begin{equation}
	|-,n\rangle=\frac{i}{\sqrt 2}\begin{bmatrix}
		\U\varphi_n\\\ -\U^*\varphi_n
	\end{bmatrix},
\end{equation}
where $\varphi_n=(0~0~\cdots~0~1~0~0~\cdots~0)^T$ with $n\in\{1,2,\cdots,M\}$ and ``$1$'' locating at the $n$th position.

Performing a unitary transformation $U=e^{-i{\pi}/{4}}e^{i({\pi}/{4})\sigma_1}$, $U\widetilde\H(\k)U^\dagger$ is real and symmetry operators can be tranformed as $U\P\T U^\dagger=\K$ and $U\SS U^\dagger=\sigma_2$. After the unitary transformation, the valence wavefunctions $U|-,n\rangle$ are real. The transition fuction $t_{S^1}$ over the intersection $S^1$ of real valence wavefunctions can be given by
\begin{equation}
	[t_{S^1}]_{mn}=\langle-,m|_N\big{|}_{S^1}U^\dagger  U|-,n\rangle_S\big{|}_{S^1}=\mathcal [\mathcal O_{S^1}]_{mn}.
\end{equation}
Thus, we know the transition function $t_{S^1}$ of real valence wavefunctions equals to the gauge transformation $\mathcal O_{S^1}$. It leads to the equivalence of two $2$D topological invariants.

\section{The Surface Effective Theories and Corresponding Projectors}\label{C}
In this section, we analytically derive surface effective theories for the Dirac model in the main text. As we shall see, the effective theories can be derived from the bulk Dirac model by applying the corresponding projectors. The surface effective theories serve as the starting point for our derivation of the effective theories for the higher-order boundaries.

We start with the upper surface [see Fig.~\ref{real3D}], while other surfaces can be treated similarly.  First, we apply the inverse Fourier transform for $k_z$ to get the first quantized Hamiltonian with the $z$-dimension in real space,
\begin{equation}\label{hh}
	\begin{split}
		\H_0=\sin& k_x\Gamma^1+\sin k_y\Gamma^2+\frac{1}{2i}(S_z-S_z^\dagger)\Gamma^3+\\
		&\left[M-\cos k_x-\cos k_y-\frac{1}{2}(S_z+S_z^\dagger)\right]\Gamma^7,
	\end{split}
\end{equation}
where $S_z$ and $S_z^\dagger$ are the forward and backward translation operators along the $z$ direction, respectively. Their actions on the real space basis of the tight-binding model for the $z$-direction are given by 
\begin{equation}
	S_z|i\rangle=|i+1\rangle,\quad S_z^\dagger|i\rangle=|i-1\rangle,
\end{equation}
where integer $i$ labeling the lattice site of the $z$ direction. Accordingly, the matrices can be explicitly written as
\begin{equation}
	S_z=\begin{bmatrix}
		\ddots&\vdots&\vdots&\vdots&\vdots&\reflectbox{$\ddots$}\\
		\ddots&0&0&0&0&\cdots\\
		\cdots&1&0&0&0&\cdots\\
		\cdots&0&1&0&0&\cdots\\
		\cdots&0&0&1&0&\cdots\\
		\reflectbox{$\ddots$}&\vdots&\vdots&\vdots&\vdots&\ddots
	\end{bmatrix},\quad S_z^\dagger=\begin{bmatrix}
		\ddots&\vdots&\vdots&\vdots&\vdots&\reflectbox{$\ddots$}\\
		\ddots&0&1&0&0&\cdots\\
		\cdots&0&0&1&0&\cdots\\
		\cdots&0&0&0&1&\cdots\\
		\cdots&0&0&0&0&\cdots\\
		\reflectbox{$\ddots$}&\vdots&\vdots&\vdots&\vdots&\ddots
	\end{bmatrix}.\nonumber
\end{equation}
For the semi-infinite systems with the surface normal to the $z$ direction, with non-negative part of the $z$ axis, we have $\widehat{S}_z^\dagger|0\rangle=0$. 
More explicitly, the semi-infinite translation operators are now written as
\begin{equation}
	\widehat{S}_z=\begin{bmatrix}
		0&0&0&0&\cdots\\
		1&0&0&0&\cdots\\
		0&1&0&0&\cdots\\
		0&0&1&0&\cdots\\
		\vdots&\vdots&\vdots&\vdots&\ddots
	\end{bmatrix},\quad \widehat{S}_z^\dagger=\begin{bmatrix}
		0&1&0&0&\cdots\\
		0&0&1&0&\cdots\\
		0&0&0&1&\cdots\\
		0&0&0&0&\cdots\\
		\vdots&\vdots&\vdots&\vdots&\ddots
	\end{bmatrix}.\nonumber
\end{equation}
Adopting the \emph{Ans$\ddot{a}$tze}
\begin{equation}
	|\psi(k_x,k_y)\rangle=\sum_{i=0}^\infty\lambda^i\ket{i}\otimes\ket{\xi(k_x,k_y)}
\end{equation}
with $|\lambda|<1$ for the surface states, we solve the Schr\"{o}dinger equation of Eq.~\eqref{hh}. In the bulk with $i\ge 1$, the Schr\"{o}dinger equation gives
\begin{widetext}
\begin{equation}
	\begin{split}
		\bigg[\sin k_x\Gamma^{1}+\sin k_y\Gamma^2+\frac{1}{2i}(\lambda^{-1}-\lambda)\Gamma^3
		+\left(M-\cos k_x-\cos k_y-\frac{1}{2}(\lambda^{-1}+\lambda)\right)\Gamma^7\bigg]\ket{\xi}=\mathcal{E}\ket{\xi}.\label{i>0}
	\end{split}
\end{equation}
Restricting to the surface layer with $i= 0$, the Schr\"{o}dinger equation leads to
\begin{equation}
	\begin{split}
		\bigg[\sin k_x\Gamma^{1}+\sin k_y\Gamma^2-\frac{1}{2i}\lambda\Gamma^3
		+\left(M-\cos k_x-\cos k_y-\frac{1}{2}\lambda\right)\Gamma^7\bigg]\ket{\xi}=\mathcal{E}\ket{\xi}\label{i=0}.
	\end{split}
\end{equation}
\end{widetext}
The difference of Eqs.~\eqref{i>0} and \eqref{i=0} gives
\begin{equation}
	i\Gamma^3\Gamma^7\ket{\xi}=\ket{\xi},\label{projector-kernel}
\end{equation}
which implies the boundary state is the eigenstate of $ i\Gamma^3\Gamma^7 $ with the eigenvalue as $ 1 $. Then, the projector for this state can be constructed as
\begin{equation}
	\Pi_1^U=\frac{1}{2}(1+i\Gamma^3\Gamma^7),
\end{equation}
where the ``1'' in $\Pi_1^U$ means the first-order projector.
Applying the projector to Eq.~\eqref{i=0}, we have
\begin{equation}
	(\sin k_x\Gamma^{1}+\sin k_y\Gamma^2)\ket{\xi}=\mathcal{E}\ket{\xi},\label{projector-does}
\end{equation}
The difference of Eqs.~\eqref{projector-does} and \eqref{i=0}, together with Eq.~\eqref{projector-kernel}, gives
\begin{equation}
	\lambda=M-\cos k_x-\cos k_y.\label{lambda}
\end{equation}
The effiective Hamiltonian for the boundary state is just
\begin{equation}
	\H^U_{eff}(\k)=\Pi_1^U\H\Pi_1^U=(\sin k_x\Gamma^{1}+\sin k_y\Gamma^2)\Pi_1^U.
\end{equation}
Similarly, we can obtain the projector of each surface
\begin{eqnarray}
	\Pi_1^{F,R,U}=\frac{1}{2}(1+ i\Gamma^{1,2,3}\Gamma^7),\\
	\Pi_1^{B,L,D}=\frac{1}{2}(1- i\Gamma^{1,2,3}\Gamma^7).
\end{eqnarray}
The total Hamiltonian is
\begin{equation}
	\H=\H_0+\Delta\H_1+\Delta\H_2,
\end{equation}
where $\Delta \H_1$ and $\Delta \H_2$ are given in the Eq.\eqref{pert}. Then, the surface Hamiltonians $\H^{\zeta}=\Pi_1^{\zeta}\H\Pi_1^{\zeta}$ are obtained by corresponding projectors. It is easy to verify that all other symmetry-preserving perturbations vanish after the projecting. It is worth noting that $\P\T$ symmetry connects the projectors of the opposite surfaces, namely, $\PP\TT\Pi_1^{F,R,U}=\Pi_1^{B,L,D}$. Meanwhile, the surface Hamiltonians of opposite surfaces are related by $\P\T$ symmetry as a natural consequence. 

\section{Theoretical Approach to Solving Second-Order and Third-Order Boundary States}\label{D}

\begin{figure}[t]
	\includegraphics[width=3.3in]{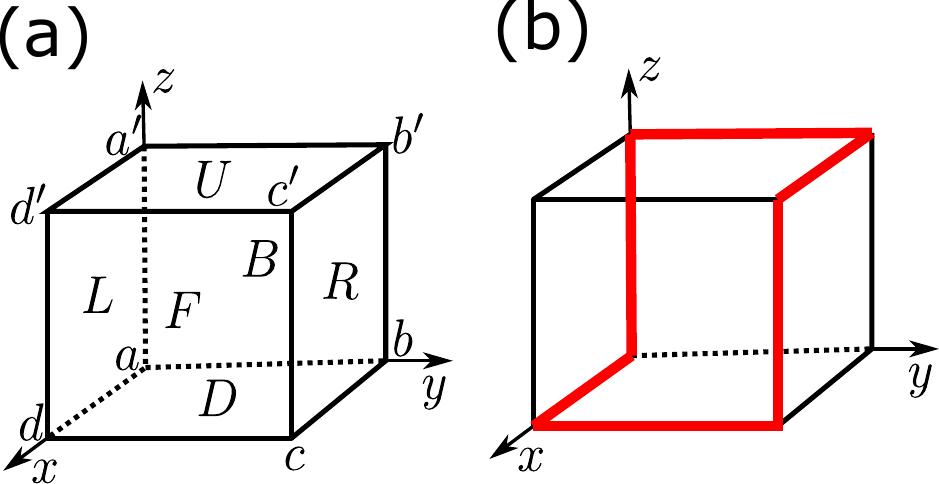}
	\caption{(a) Real space in 3D. (b) The distribution of helical states with $\alpha_1<0$ and $\alpha_2>0$ in Eqs. (\ref{alpha1}, \ref{alpha2}).}\label{real3D}
\end{figure}

We first give the results for appearing second-order helical states with respect to first-type perturbation $\Delta\H_1$.	Consider a cubic shape sample as shown in Fig.~\ref{real3D}(a). With the first-type perturbation $\Delta\H_1$ added, the gapless surface states perpendicular to $x/y/z$-axis require that $\lambda_{(1/2/3)4}=\lambda_{(1/2/3)5}=0$, and the helical edges parallel to $x/y/z$-axis require that 
\begin{eqnarray}
	x:&&~(\lambda_{24},\lambda_{25})=\alpha_1(\lambda_{34},\lambda_{35}),\label{alpha1}\\
	y:&&~(\lambda_{14},\lambda_{15})=\alpha_2(\lambda_{34},\lambda_{35}),\label{alpha2}\\
	z:&&~(\lambda_{14},\lambda_{15})=\alpha_3(\lambda_{24},\lambda_{25}),\label{alpha3}
\end{eqnarray}
with $\alpha_i$ is arbitrary nonzero real number. If $\alpha_{1/2/3}>0 ~(\alpha_{1/2/3}<0)$, the helical edge states distribute on a pair of diagonal (off-diagonal) edges parallel to $x/y/z$-axis. The diagonal edges parallel to $x$-axis: $a'd'$ and $bc$; $y$-axis: $a'b'$ and $dc$; $z$-axis: $bb'$ and $dd'$. The off-diagonal edges parallel to $x$-axis: $ad$ and $b'c'$; $y$-axis: $ab$ and $d'c'$; $z$-axis: $aa'$ and $cc'$. Note that edges in each pair are related by $\PP\TT$ symmetry.
If none of the three equations above is satisfied, there exist zero-mode corner states, instead of the helical edge states, at a pair of antipodal vertexes in the cube. Next, we present the proof for Eq. \eqref{alpha1} as an example, and the proofs for other two equations are similar.

Consider the edge $ad$ ($b'c'$ is related to $ad$ by $\mathcal{PT}$) as shown in Fig. \ref{real3D}. The effective Hamiltonians for the two relevant surfaces $D$ and $L$ are given by
\begin{equation}
	\begin{split}
		\H_1^D&=\left[\sin k_x\Gamma^1+\sin k_y\Gamma^2+i\Gamma^3(\lambda_{34}\Gamma^4+\lambda_{35}\Gamma^5)\Gamma^7\right]\Pi_1^D,\\
		\H_1^L&=\left[\sin k_x\Gamma^1+\sin k_z\Gamma^3+i\Gamma^2(\lambda_{24}\Gamma^4+\lambda_{25}\Gamma^5)\Gamma^7\right]\Pi_1^L.
	\end{split}
\end{equation}
To obtain the helical hinge states on $ad$, open boundaries normal to the $y$-axis for the effective Hamiltonian $\H_1^ D$ and boundaries normal to the $z$-axis for $\H_1^ L$. And simplify the surface effective Hamiltonians by replacing $\sin k_{y,z}$ with $k_{y,z}$ or $-i\partial_{y,z}$. If there exist helical hinge states along the $x$-axis, they must be localized at and decay exponentially away from the hinge. Therefore, we can adopt the \emph{Ans$\ddot{a}$tze} for the helical states with respect to $k_x$ on the two surfaces as
\begin{equation}
	\Psi^D(y,k_x)=\Psi^D_0(k_x)e^{-\lambda_yy},~\Psi^L(z,k_x)=\Psi^L_0(k_x)e^{-\lambda_zz},\label{solutions}
\end{equation}
where the decay rates $\lambda_{x,y}>0$ and $\Psi^{D/L}$ satisfy that $\H_1^{D}\Psi^{D}(y,k_x)=\sin k_x\Gamma^1\Pi_1^{D}\Psi^{D}(y,k_x)$, $\H_1^{L}\Psi^{L}(y,k_x)=\sin k_x\Gamma^1\Pi_1^{L}\Psi^{L}(y,k_x)$. Then, the helical states can be solved by the equations
\begin{equation}
	\begin{split}
		\Gamma^2\Gamma^3(\lambda_{34}\Gamma^4+\lambda_{35}\Gamma^5)\Gamma^7\Psi_0^D&=-\lambda_y\Psi_0^D,\\
		-\Gamma^2\Gamma^3(\lambda_{24}\Gamma^4+\lambda_{25}\Gamma^5)\Gamma^7\Psi_0^L&=-\lambda_z\Psi_0^L,
	\end{split}\label{solution}
\end{equation}
with the boundary continuous condition $\Psi_0^D=\Psi_0^L$ that connects $\Psi^{D}$ and $\Psi^{L}$ at hinge $ad$. The decay rates are given as $\lambda_y=\sqrt{\lambda^2_{34}+\lambda_{35}^2}$ and $\lambda_z=\sqrt{\lambda^2_{24}+\lambda_{25}^2}$. 
As a result, $\Psi_0$ is simultaneously the eigenstate with eigenvalue $-1$ for operators $\Lambda_1=\Gamma^2\Gamma^3(\lambda_{34}\Gamma^4+\lambda_{35}\Gamma^5)\Gamma^7/\lambda_y$ and $\Lambda_2=-\Gamma^2\Gamma^3(\lambda_{24}\Gamma^4+\lambda_{25}\Gamma^5)\Gamma^7/\lambda_z$. According to the anticommutation relations of $\Gamma$ matrices, $\Lambda_1,\Lambda_2$ have the same set of eigenstates $\{\Psi_0,\Gamma^2\Psi_0,\Gamma^3\Psi_0,\Gamma^7\Psi_0,\Gamma^2\Gamma^3\Psi_0,\Gamma^2\Gamma^7\Psi_0,\Gamma^3\Gamma^7\Psi_0,\\ \Gamma^2\Gamma^3\Gamma^7\Psi_0\}$, so they must commute with each other, resulting in $\lambda_{34}\lambda_{25}=\lambda_{24}\lambda_{35}$, or, equivalently, Eq.~\eqref{alpha1}. Substituting Eq.~\eqref{alpha1} into $\Lambda_1$ and utilizing again the fact that $\Lambda_1$ and $\Lambda_2$ have the same eigenstate $\Psi_0$ with the same eigenvalue $-1$, we find that $\alpha_1=-\lambda_z/\lambda_y<0$. 
Therefore, we conclude that once the Eq.~(\ref{alpha1}) holds with $\alpha_1<0$ and $(\lambda_{24},\lambda_{25})\neq 0$, there are $\mathcal{PT}$-related helical modes along $ad$ and $b'c'$.

Now we proceed to discussion the theoretical analysis for third-order topological phases with respect to second-type perturbation $\Delta\H_2$. Although
$\Delta\H_2$ cannot gap out the gapless surfaces, it can gap out the helical hinge states. 
Let both two Eqs. (\ref{alpha1},\ref{alpha2})  hold with $\alpha_1<0$ and $\alpha_2>0$, which implies the all three equations hold and $\alpha_3<0$. In this case, the helical states will distribute along the six hinges, namely, $aa'$-$a'b'$-$b'c'$-$c'c$-$cd$-$da$, as shown in Fig. \ref{real3D}(b). With $\Delta\H_2$ added, the effective Hamiltonians for the two relevant surfaces can be simplified as
\begin{equation}
	\begin{split}
		\H_2^D&=\H_1^D+i(\eta_1\Gamma^1+\eta_2\Gamma^2)\Gamma^6\Pi_1^D,\\
		\H_2^L&=\H_1^L+i(\eta_1\Gamma^1+\eta_3\Gamma^3)\Gamma^6\Pi_1^L.
	\end{split}\label{Second-order-eff}
\end{equation}
From Eqs. \eqref{solution}, we know that the helical states $\Psi_0^{D/L}$ along the $x$ axis is the eigenstates of  $\Gamma^2\Gamma^3(\widetilde{\lambda}_{34}\Gamma^4+\widetilde{\lambda}_{35}\Gamma^5)\Gamma^7$, where $\widetilde{\lambda}_{34,35}=\lambda_{34,35}/\sqrt{\lambda_{34}^2+\lambda_{35}^2}$. Hence, we can natually construct a new projector, namely,
\begin{equation}
	\begin{split}
		\Pi_2^{ad}&=\frac{1}{2}\big(1+\Gamma^2\Gamma^3(\widetilde{\lambda}_{34}\Gamma^4+\widetilde{\lambda}_{35}\Gamma^5)\Gamma^7\big),\\
		&(\Pi_2^{ad})^2=\Pi_2^{ad},\quad [\Pi_2^{ad},\Pi_1^{D/L}]=0,\label{second-projector}
	\end{split}
\end{equation}
where the $``2"$ in $\Pi_2^{ad}$ means the second-order projector.
The projector $\Pi_2^{ad}$ can project surface Hamiltonian of $D$ and $L$ to the hinge $ad$. Following the same argument, we can obtain another projector $\Pi_2^{aa'}$ which can project surface Hamiltonian to hinge $aa'$.
\begin{equation}
	\begin{split}
		\Pi_2^{aa'}&=\frac{1}{2}\big(1+\Gamma^1\Gamma^2(\widetilde{\lambda}_{34}\Gamma^4+\widetilde{\lambda}_{35}\Gamma^5)\Gamma^7\big),\\
		(&\Pi_2^{aa'})^2=\Pi_2^{aa'},\quad [\Pi_2^{aa'},\Pi_1^{L/B}]=0.\label{second-projector_P2}
	\end{split}
\end{equation}
Applying the projectors (\ref{second-projector}, \ref{second-projector_P2}) to the surface Hamiltonians, we can obtain the effective Hamiltonians for the edges $ad,aa'$,
\begin{equation}
	\begin{split}
		\H_{ad}(k_x)&=\Gamma^1(k_x+i\eta_1\Gamma^6)\Pi_2^{ad}\Pi_1^{D/L},\\
		\H_{aa'}(k_z)&=\Gamma^3(k_z+i\eta_2\Gamma^6)\Pi_2^{aa'}\Pi_1^{L/B}.
	\end{split}\label{edgeH}
\end{equation}
Since $i\Gamma^{1,3}\Gamma^6$ anticommutes with $\Gamma^{1,3}$, the helical modes along $ad$ and $aa'$ are gapped out by $i\eta_{1,3}\Gamma^{1,3}\Gamma^6$. To solve the zero modes at corner $a$, we replace $k_{x,z}$ by $-i\partial_{x,z}$ in Eqs. \eqref{edgeH}, which leads to
\begin{equation}
	\begin{split}
		\partial_x|\psi(x)\rangle=\eta_1\Gamma^6|\psi(x)\rangle,\quad \partial_z|\phi(z)\rangle=\eta_3\Gamma^6|\psi(z)\rangle.
	\end{split}
\end{equation}
Similarly, if there exists a zero mode at corner $a$, the state must be localized at and decays exponentially away from the corner. Thus we can adopt the  \emph{Ans$\ddot{a}$tze} for the corner zero mode,
\begin{equation}
	|\psi(x)\rangle=\psi_0e^{-|\eta_1|x},\quad |\phi(z)\rangle=\phi_0e^{-|\eta_3|z}.
\end{equation}
Then, we have
\begin{equation}
	\mathrm{sgn}(\eta_1)\psi_0=\Gamma^6\psi_0,\quad \mathrm{sgn}(\eta_3)\phi_0=\Gamma^6\phi_0.\label{corner}
\end{equation}
The continuity condition requires that $\psi_0=\phi_0$ at  corner $a$. Thus, we obtain sgn($\eta_1$)=sgn($\eta_3$). Finally, we conclude that when the first-type perturbations satisfy $\alpha_1<0$ and $\alpha_2>0$ in Eqs. (\ref{alpha1}, \ref{alpha2}), the zero mode appearing at corner $a$ requires that the second-type perturbations satisfies $\mathrm{sgn}(\eta_1)=\mathrm{sgn}(\eta_3)$. 

In fact, from Eq. \eqref{corner}, we can know that $\psi_0=\phi_0$ is the eigenstate of $\Gamma^6$ with the eigenvalue as sgn($\eta_1$)$=$sgn($\eta_3$). Thus we can construct a third-order projector as
\begin{equation}
	\Pi_3^a=\frac{1}{2}(1+\mathrm{sgn}(\eta_1)\Gamma^6).\label{third-projector}
\end{equation}
Applying the third-order projector \eqref{third-projector} to Eqs. \eqref{edgeH}, we can naturally obtain the zero-energy corner Hamiltonian, which corresponds to a corner zero mode at $a$.

\section{The derivation details for the parity condition}\label{E}
Cubic lattice is applied to our Dirac model. For the integrity of each unit cell, sample geometry is limited to a structure composed of several cuboids. Therefore, all surfaces of the sample can be devided into six categories: U, D, F, B, L and R, which is shown in the FIG.~\ref{real3D}(a). Also, each corner must be the common point of three surfaces: U/D, F/B and L/R. For example, point $a$ in FIG.~\ref{real3D}(a) is the intersection point of surfaces D, B, and L. No matter how the sample geometry changes, there are always eight kinds of corners, which can be labled as \{U/D, F/B, L/R\}. \\

In the above section ``\emph{Theoretical Approach to Solving Second-Order and Third-Order Boundary States}'', we have analyzed the position of zero-mode corners under different perturbation conditions. When the Hamiltonian $\H_0+\Delta\H$ are fixed, the existence of zero-mode corner states at a given corner only depends on which surfaces intersects this corner, i.e., the kind of this corner.\\

We have known that there are $1$ or $3$  pairs of $\PP\TT$ related zero mode corners in the cube sample. In other words, there are $2$ or $6$ kinds of zero-mode corners in total eight kinds of corners no matter how the sample geometry changes. To meet the requirement of  $\PP\TT$ symmetry, sample geometry must be inversion symmetric. In the inversion symmetric sample, each kind of corner must appear an odd number of times. Suppose the occurrence number of these eight kinds of corners are 
$$\{n_1,~ n_2,~n_3,~n_4,~n_1,~n_2,~n_3,~n_4\},~~~ n_i \text{ is odd }.$$
Two identical $n_i$ is because two kinds of corners are related one to one by inversion symmetry. The number of zero-mode corner pair is $n$. We can know that
\begin{eqnarray}
	&n=n_i,\label{pairnumber1}\\ 
	&\text{or }n=n_1+n_2+n_3+n_4-n_i,\label{pairnumber2}
\end{eqnarray}
where Eq.~(\ref{pairnumber1}) and Eq.~(\ref{pairnumber2}) correspond to $2$ and $6$ kinds of zero-mode corners, respectively. $i=1,2,3,4$ depends on the perturbation terms. Therefore, $n$ is always odd.
\begin{figure}[t]
	\includegraphics[width=2in]{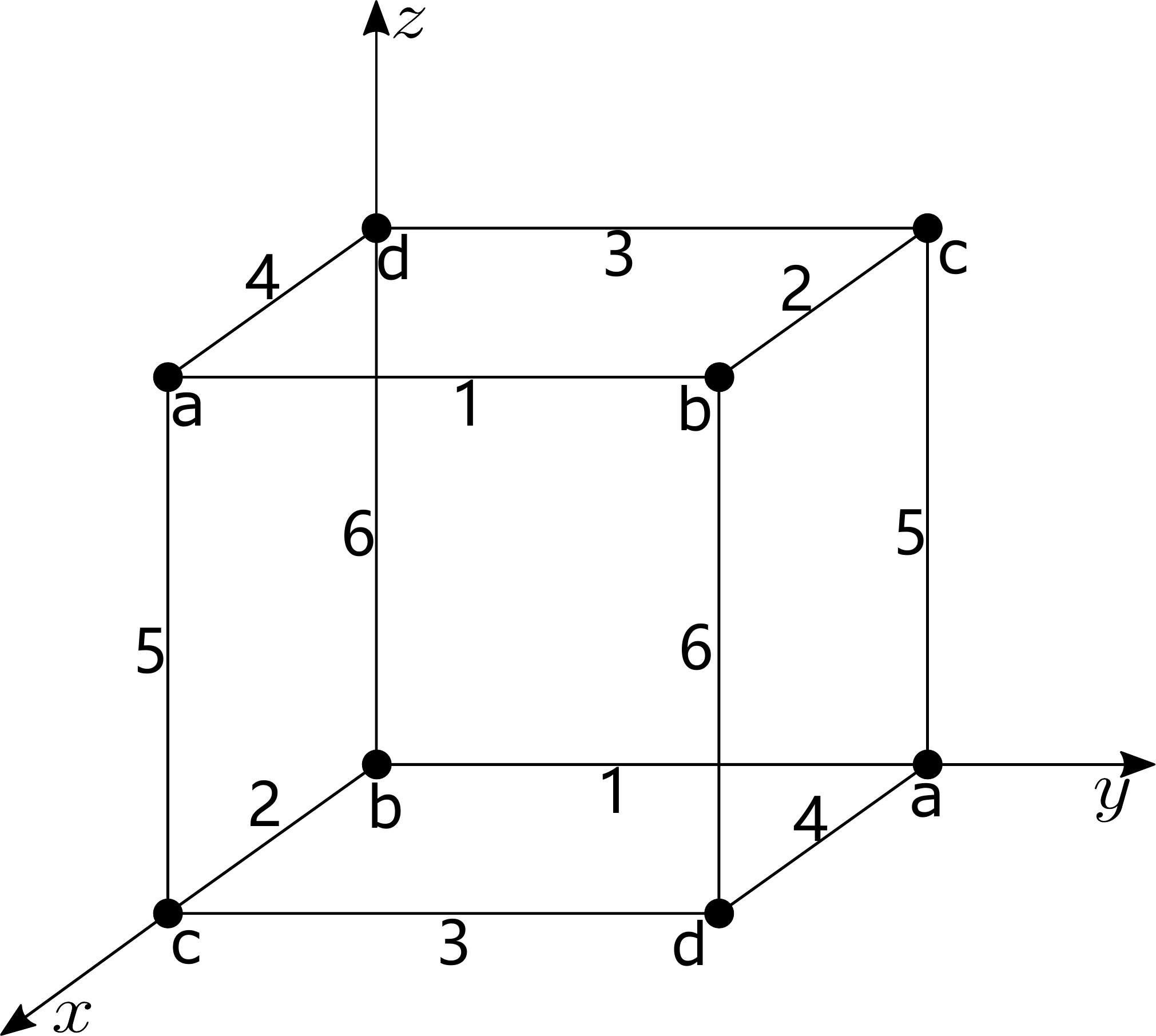}
	\caption{The labels for gapless edges and corners with zero modes.}
	\label{Naming_Scheme}
\end{figure}

\section{Naming Scheme of boundary states}\label{F}
Boundary regions with in-gap states are paired by $\PP\TT$ symmetry. Hence, gapless surfaces are labeled by $S_{x,y,z}$, where the subscript denotes the normal direction of the gapless surfaces. Gapless edges are labeled by $E_{i}$ with $i=1,2,\cdots,6$ as shown in Fig. \ref{Naming_Scheme}.  Corners are denoted by $C_{\alpha}$ with $\alpha=a,b,c,d$ as shown in Fig. \ref{Naming_Scheme}.

\section{The dimensions of all possible phases and the phase transitions between them}\label{G}
\begin{figure}[t]
	\includegraphics[width=3.3in]{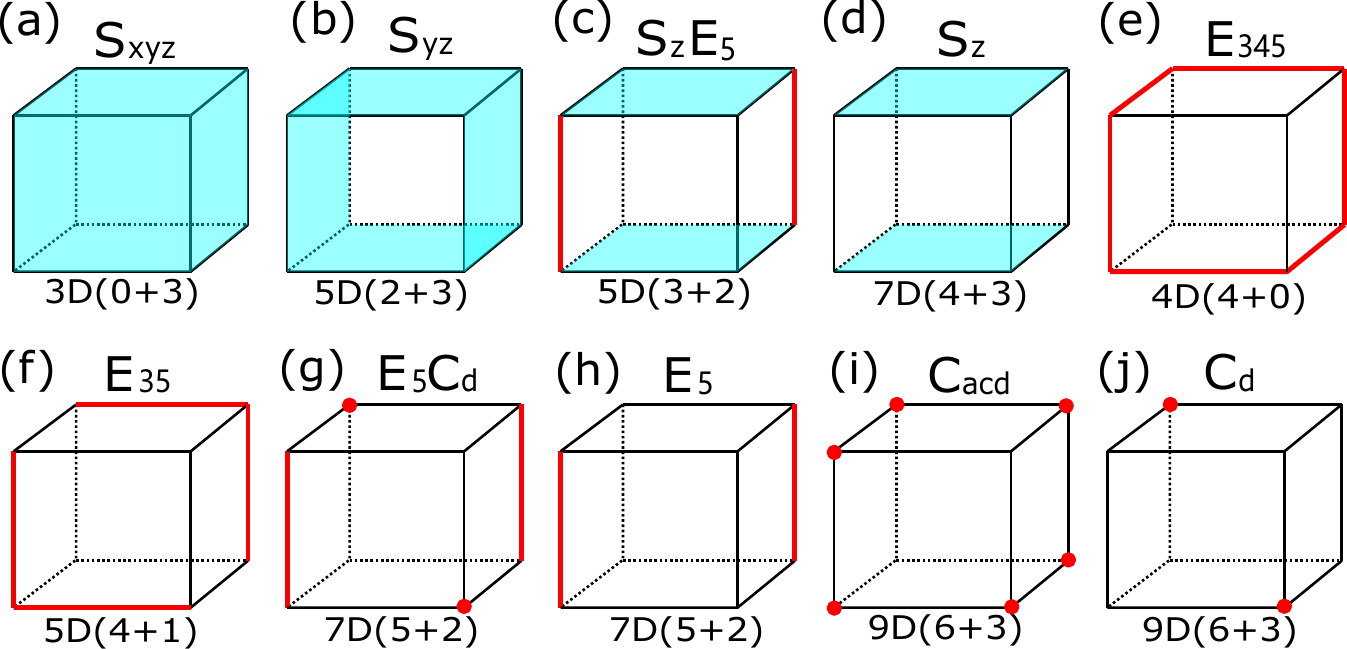}
	\caption{All possible topological-phase configurations and their highest cellular dimensions.}
	\label{hierarchy}
\end{figure}
\begin{figure*}[t]
	\includegraphics[scale=0.55]{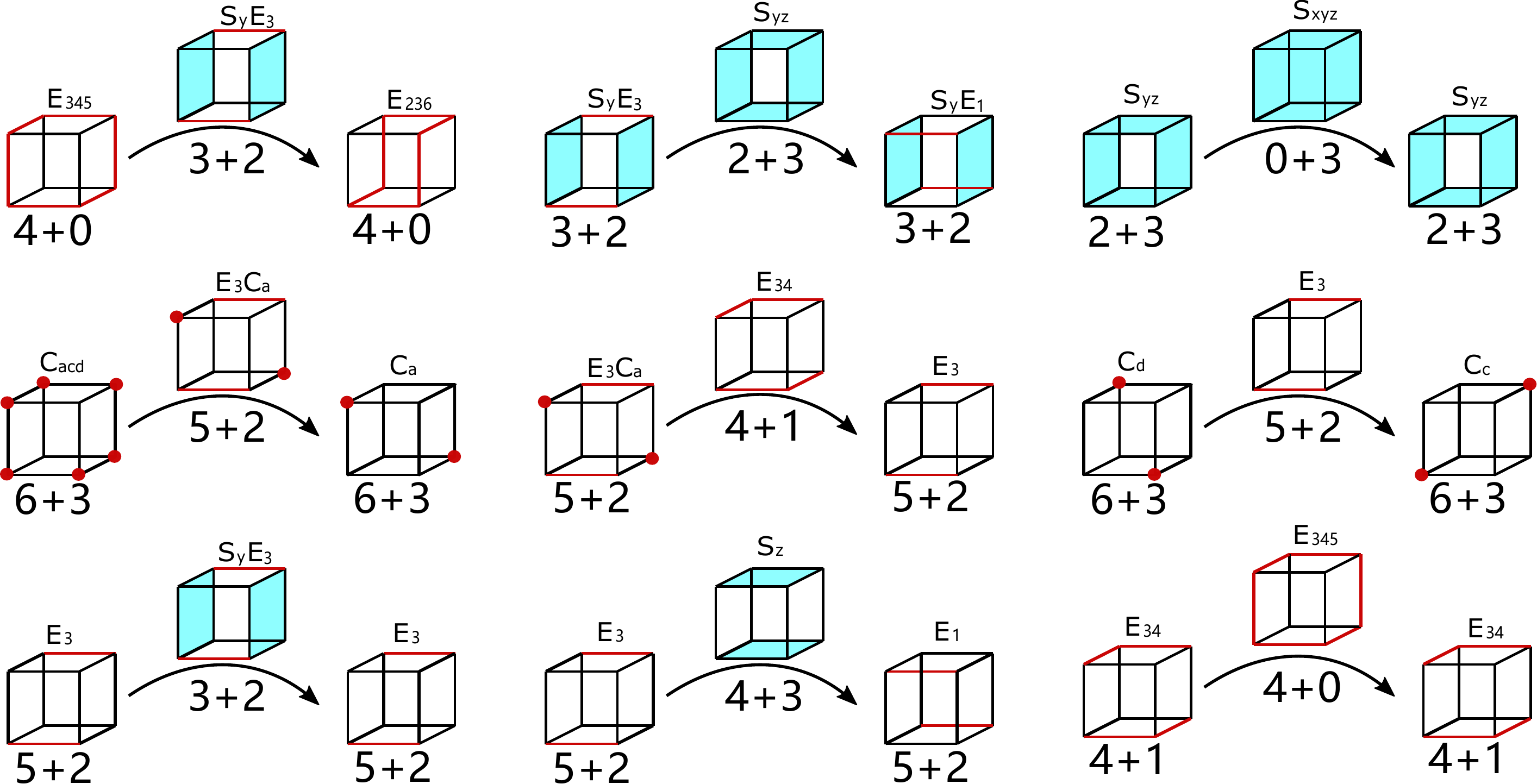}
	\caption{The typical phase transitions between the same kind of configurations as the parameters change continuously.}
	\label{phase-transitions}
\end{figure*}
The all possible topological-phase configurations and their cellular dimension are elucidated in Fig. \ref{hierarchy}. In bracket $(D_1+D_2)$, $D_1$ represents the dimension of first-type perturbations and $D_2$ the dimension of second-type perturbations. Based on the fact that the first-type perturbations can generally gap out the gapless surface states while the second-type perturbations can gap out the gapless hinge states, we have two equations of cellular dimensions with respect to the number of  first-order or second-order gapless boundaries: 
\begin{eqnarray}
	D_1&=&6-N_S-N_L/2+[N_L/6],\label{D1}\\
	D_2&=&3-N_L/2,\label{D2}
\end{eqnarray}
where $N_S$ is the number of first-order gapless boundaries, namely, gapless surfaces, $N_L$ the number of second-order gapless boundaries, namely, gapless hinges, and ``[~]'' is the floor function. Due to the $\P\T$ symmetry, the gapless boundaries must appear or disappear in inversion-related pairs.  Among the six first-type perturbation terms, $\lambda_{i4},\lambda_{j5},i,j=1,2,3.$, each pair $(\lambda_{i4},\lambda_{i5})$ with $i=1,2,3$ can gap the same pair of gapless surfaces. Thus, if the number of first-order gapless boundaries increases by a pair, the freedom degree of first-order perturbations will decrease by $2$, which corresponds to $N_S$ term in Eq. \eqref{D1}.
From Eqs.~(\ref{alpha1},\ref{alpha2},\ref{alpha3}), we find that if the number of second-order boundaries increases by a pair, one equation of the three must hold. This imposes a restriction to the six freedom degrees of freedom for the first-type perturbations and therefore decreases one cellular dimension, corresponding to $N_L/2$ term in Eq.~\eqref{D1}. However, it is noted that only two equations among the three Eqs.~(\ref{alpha1},\ref{alpha2},\ref{alpha3}) are independent, namely any two of them implies the third. This justifies the correction term $[N_L/6]$ in Eq.~\eqref{D1}, since two pairs of gapless hinges and three pairs of gapless hinges decrease the same degrees of freedom for the first-type perturbations.	 	
For $D_2$, it is found from Eq.~\eqref{edgeH} that each second-type perturbation term can gap out an inversion pair of gapless hinges parallel to a coordinate axis. More specifically, $\eta_1\Gamma^i\Gamma^6$ gaps out the gapless hinges parallel to the $i$-axis, with $i=1,2,3$.   Hence, if the number of second-order gapless boundaries (gapless hinges) is increased by a pair, the degrees of freedom for the second-type perturbations will be decreased by one, which corresponds to the term $N_L/2$ in Eq.~\eqref{D2}.

The typical phase transitions between the same kind of configurations but with different boundary distributions are illustrated in Fig.\ref{phase-transitions}. In the process of phase transitions, it is found that the critical phases always have a lower or equal dimension of first-type perturbations. However, the cellular dimension for the second-type perturbations is completely determined by the number of gapless hinges, obeying the Eq.~\eqref{D2}.\\

\section{Numerical Results}\label{H}
In this section, we present the numerical results for each phase in Fig.~\ref{hierarchy}. The numerical results shown in FIG.~\ref{hierarchy}(a)-(j) correspond to boundary-state configurations in Fig.\ref{tena}-\ref{tenj}, respectively. We list the corresponding boundary-state configuration for each figure below.

The numerical results for ``$S_{xyz}$'' are shown in FIG. \ref{tena}, which corresponds to the first-order topological phases with six gapless surfaces.  

The numerical results for ``$S_{yz}$'' are shown in FIG. \ref{tenb}, which corresponds to the first-order topological phases with four gapless surfaces parallel to $x$-axis.

The numerical results for ``$S_zE_5$'' are shown in FIG. \ref{tenc}, which corresponds to the mix-order topological phases with two $x$-$y$ gapless surfaces connected by two $\PP\TT$-related helical $z$-edges. 

The numerical results for ``$S_z$'' are shown in FIG. \ref{tend}, which corresponds to the first-order topological phases with a pair of $x$-$y$ surfaces.

The numerical results for ``$E_{345}$'' are shown in FIG. \ref{tene}, which corresponds to the second-order topological phases with three pairs of $\P\T$-related helical edges.  

The numerical results for ``$E_{35}$'' are shown in FIG. \ref{tenf}, which corresponds to the second-order topological phases with two pairs of $\P\T$-related helical edges. 

The numerical results for ``$E_{5}C_d$'' are shown in FIG. \ref{teng}, which corresponds to the mix-order topological phases with two $\P\T$-related $z$-edges and a pair of antipodal zero modes $d$-corners.

The numerical results for ``$E_{5}$'' are shown in FIG. \ref{tenh}, which corresponds to the second-order topological phases with two $\P\T$-related helical $z$-edges.

The numerical results for ``$C_{acd}$'' are shown in FIG. \ref{teni}, which corresponds to the third-order topological phases with three pairs of antipodal zero modes corners.

The numerical results for ``$C_{d}$'' are shown in FIG. \ref{tenj}, which corresponds to the third-order topological phases with a pair of antipodal zero modes corners.

The numerical results for FIG.~2 in the maintext are shown in FIG. \ref{simulation}. Odd pairs of zero-mode corner states are verified in different sample geometry.

\begin{figure*}[h]
	\includegraphics[scale=0.9]{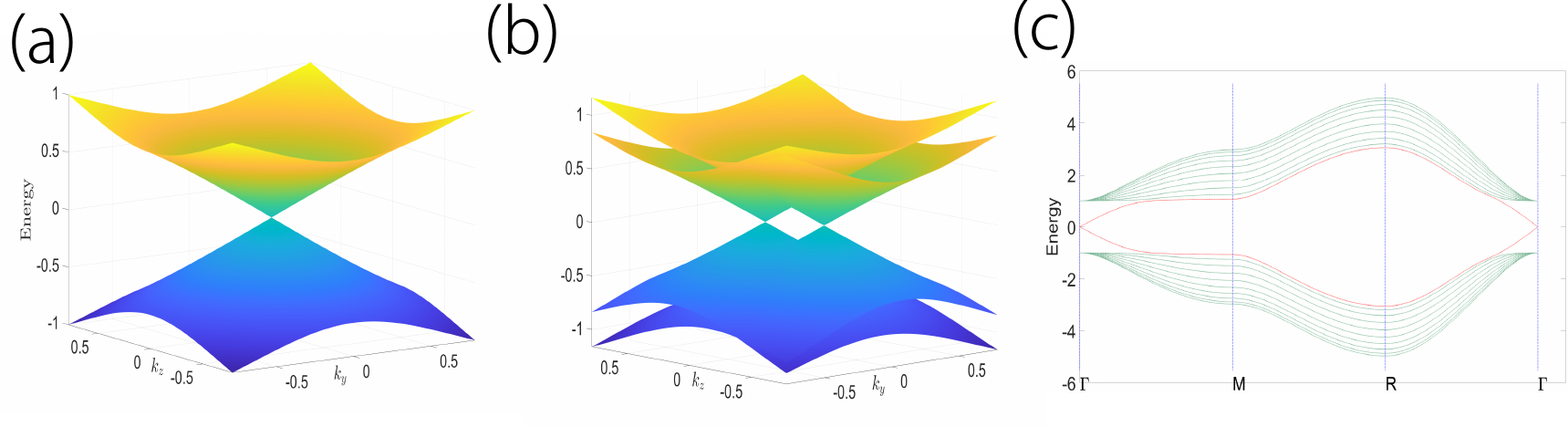}
	\caption{The numerical results for Fig.~\ref{hierarchy}(a). The cellular dimension is $3$D($0+3$). (a) shows the Dirac cone on the surfaces parallel to the $yz$ plane, plotted by the boundary effective Hamiltonian without perturbations.  The other four surfaces are also gapless. (b) shows the energy spectrum of the boundary effective Hamiltonian with perturbations $i(\eta_1\Gamma^1+\eta_2\Gamma^2+\eta_3\Gamma^3)\Gamma^6$ on the surfaces parallel to the $yz$ plane. If such perturbations are added, a fourfold degenerate Dirac point will split into two twofold degenerate Dirac points. We adopt the parameters: $\eta_1=0.1,~\eta_2=0.15,~\eta_3=0.08$. (c) shows the spectrum for the high-symmetric lines $\Gamma-M-R-\Gamma$ with all perturbations absent, corresponding to (a). If the perturbations are added, each Dirac point will split into two and may deviate from the high-symmetric lines. The surface is still gapless.}\label{tena}
\end{figure*}

\begin{figure*}[h]
	\includegraphics[scale=0.7]{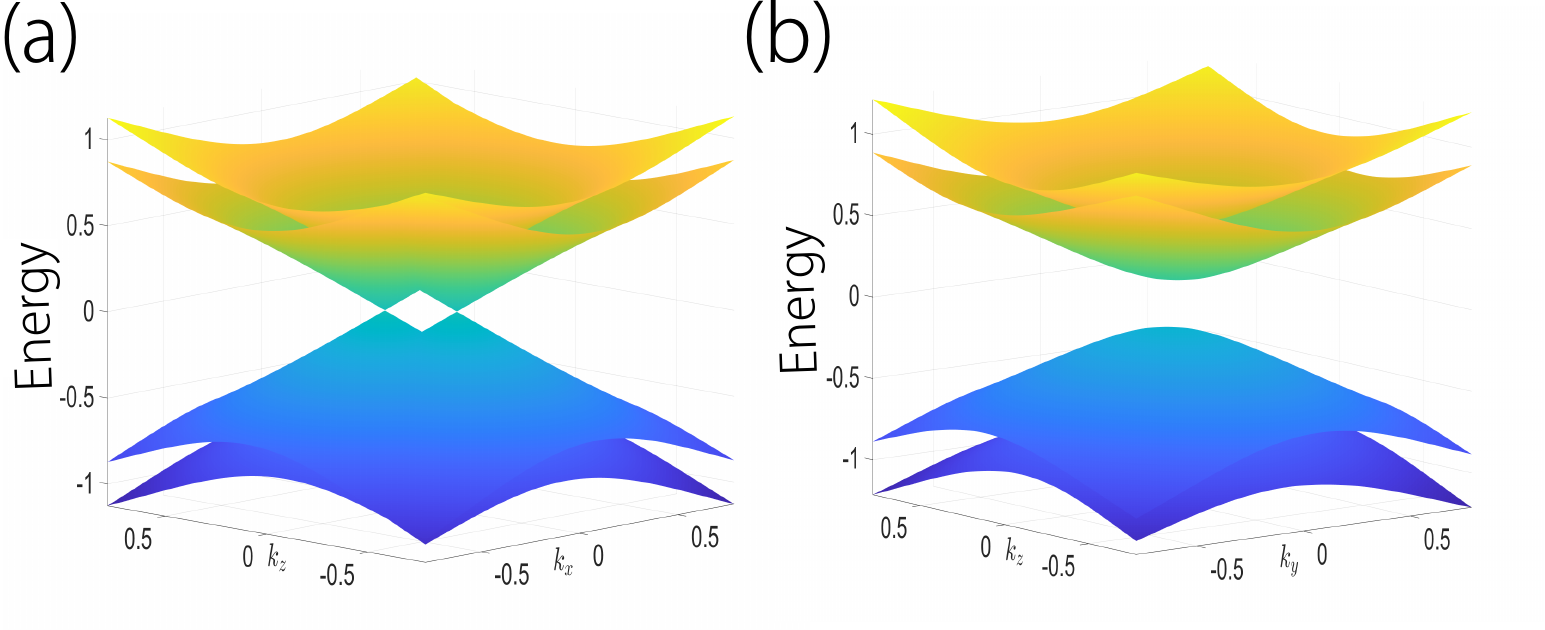}
	\caption{The numerical results for Fig.~\ref{hierarchy}(b). The cellular dimension is $5$D($2+3$). The permitted perturbations are $i\Gamma^1(\lambda_{14}\Gamma^4+\lambda_{15}\Gamma^5)\Gamma^7+i(\eta_1\Gamma^1+\eta_2\Gamma^2+\eta_3\Gamma^3)\Gamma^6$. So the parameters are set as  $\lambda_{14}=0.2,~\lambda_{15}=0.25,~\eta_1=0.1,~\eta_2=0.15,~\eta_3=0.08$. (a) shows the energy spectrum of the boundary effective Hamiltonian on the gapless surfaces parallel to the $xz$ plane. The surfaces parallel to the $xy$ plane are also gapless. (b) shows the spectrum of the boundary effective Hamiltonian on the gapped surfaces parallel to the $yz$ planes.}\label{tenb}
\end{figure*}

\begin{figure*}
	\includegraphics[scale=0.45]{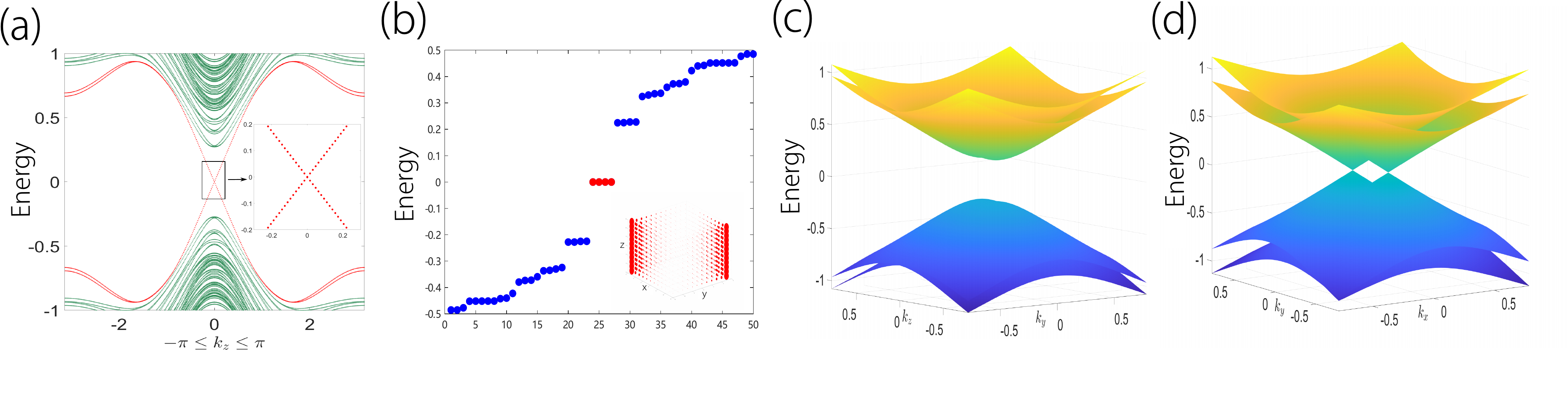}
	\caption{The numerical results for Fig.~\ref{hierarchy}(c). The cellular dimension is $5$D($3+2$). The permitted perturbations are $i\Gamma^1(\lambda_{14}\Gamma^4+\lambda_{15}\Gamma^5)\Gamma^7+i\Gamma^2(\lambda_{24}\Gamma^4+\lambda_{25}\Gamma^5)\Gamma^7+i(\eta_1\Gamma^1+\eta_2\Gamma^2)\Gamma^6$, where $(\lambda_{14},\lambda_{15})=\alpha_3(\lambda_{24},\lambda_{25})$ with $\alpha_3>0$. So the parameters are set as $(\lambda_{14},\lambda_{15})=(\lambda_{24},\lambda_{25})=(0.2,0.3),~\eta_1=0.1,~\eta_2=0.08.$ (a) shows the spectrum with respect to $k_z$ (open the boundaries normal to $x$ and $y$ axis). It is observed that there exist gapless states along the $k_z$ axis. (b) shows the energy spectrum in real space with the periodic boundary conditions along $z$ direction. The right insert shows the spatial distribution of the four in-gap states. (c) shows the spectrum of boundary effective Hamiltonians on the gapless surfaces parallel to the $xy$ plane. (d) shows the spectrum of boundary effective Hamiltonians on the gapped surfaces parallel to the $yz$ plane. The surfaces parallel to the $xz$ plane are also gapped. 
	}\label{tenc}
\end{figure*}

\begin{figure*}
	\includegraphics[scale=0.9]{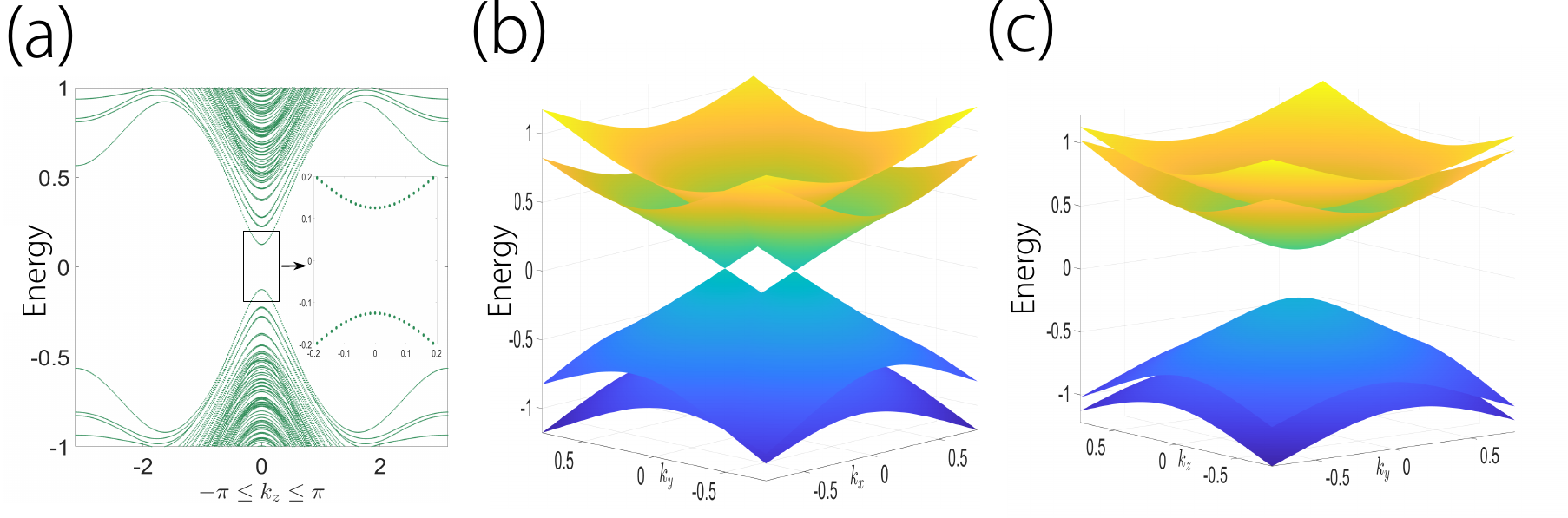}
	\caption{The numerical results for Fig.~\ref{hierarchy}(d). The cellular dimension is $7$D($4+3$). The permitted perturbations are $i\Gamma^1(\lambda_{14}\Gamma^4+\lambda_{15}\Gamma^5)\Gamma^7+i\Gamma^2(\lambda_{24}\Gamma^4+\lambda_{25}\Gamma^5)\Gamma^7+i(\eta_1\Gamma^1+\eta_2\Gamma^2+\eta_3\Gamma^3)\Gamma^6$. So the parameters are set as  $\lambda_{14}=0.2,~\lambda_{15}=0.3,~\lambda_{24}=0.2,~\lambda_{25}=0.15,~\eta_1=0.08,~\eta_2=0.1,~\eta_3=-0.12.$ (a) shows the spectrum with respect to $k_z$ (with open boundary conditions for surfaces normal to the $x$ and $y$ axis). It is observed that the surface spectrum along the $k_z$ axis is gapped out by perturbations. (b) The spectrum on the gapless surfaces parallel to the $xy$ plane. (c) The spectrum on the gapped surfaces parallel to the $yz$ plane. The surfaces parallel to the $xz$ plane are also gapped.	}\label{tend}
\end{figure*}

\begin{figure*}
	\includegraphics[scale=0.9]{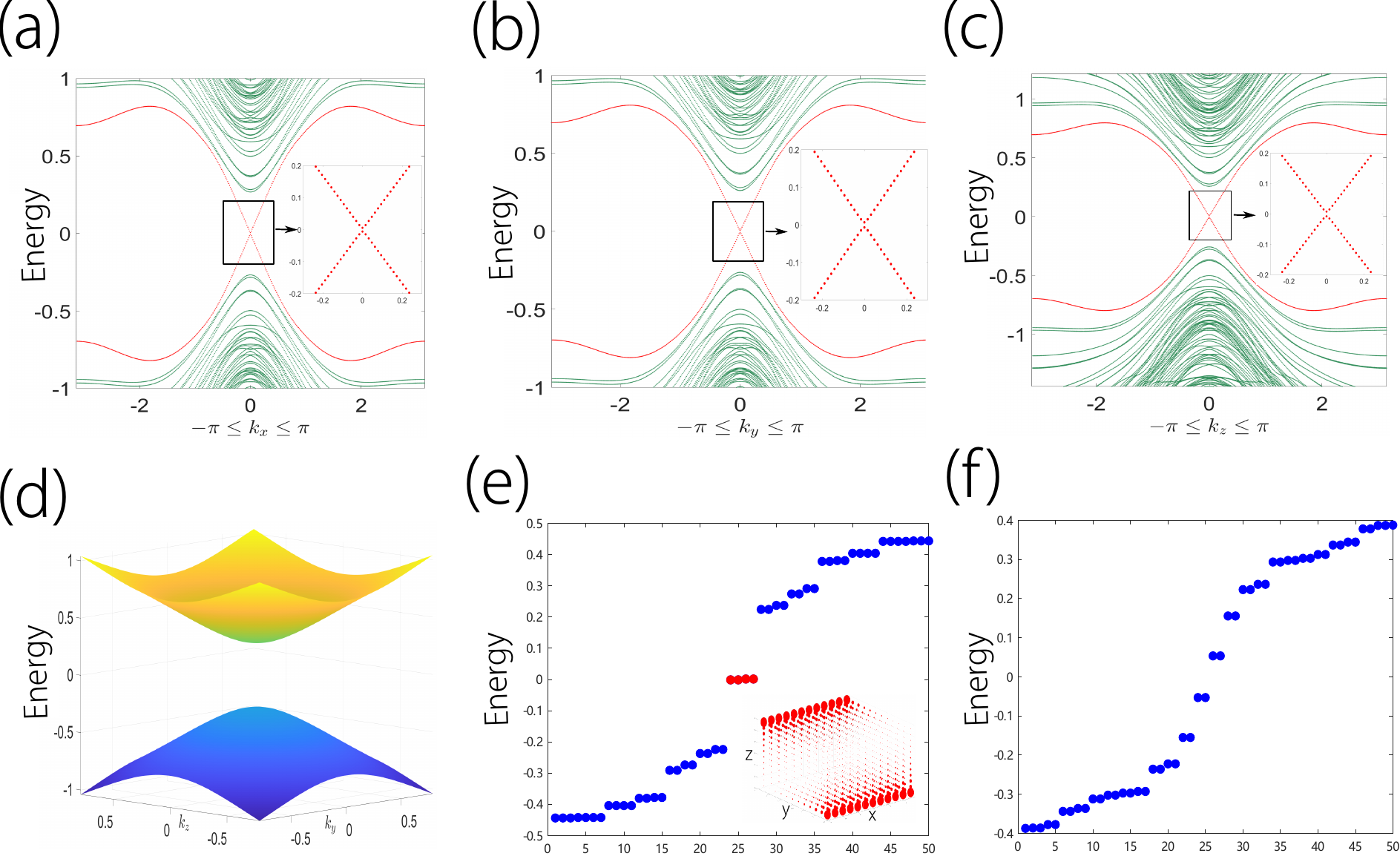}
	\caption{The numerical results for Fig.~\ref{hierarchy}(e). The cellular dimension is $4$D($4+0$). The permitted perturbations are $i\Gamma^1(\lambda_{14}\Gamma^4+\lambda_{15}\Gamma^5)\Gamma^7+i\Gamma^2(\lambda_{24}\Gamma^4+\lambda_{25}\Gamma^5)\Gamma^7+i\Gamma^3(\lambda_{34}\Gamma^4+\lambda_{35}\Gamma^5)\Gamma^7$, where $(\lambda_{24},\lambda_{25})=\alpha_1(\lambda_{34},\lambda_{35}),(\lambda_{14},\lambda_{15})=\alpha_2(\lambda_{34},\lambda_{35}),(\lambda_{14},\lambda_{15})=\alpha_3(\lambda_{24},\lambda_{25})$ with $\alpha_{1,2,3}>0$. So the parameters are set as $\lambda_{14}=\lambda_{15}=0.2,~\lambda_{24}=\lambda_{25}=0.21,~\lambda_{34}=\lambda_{35}=0.22.$ (a-c) presents the energy spectra in the $k_{x/y/z}$ momentum space with the periodic boundary conditions along the $x/y/z$ axis, respectively.  (d) presents the spectrum on the gapped surfaces parallel to the $yz$ plane. Similarly, the other four surfaces are also gapped. (e) shows the energy spectrum in real space with the periodic boundary conditions along $x$ direction. The right insert denotes the sptial distribution of the four in-gap states. The energy spectra in real space with the periodic boundary conditions along $y$ or $z$ direction are similarly. And the composition of them just gives Fig. \ref{hierarchy}(e). (f) shows the energy spectrum in real space with none periodic direction.}\label{tene}
\end{figure*}

\begin{figure*}[h]
	\includegraphics[width=7in]{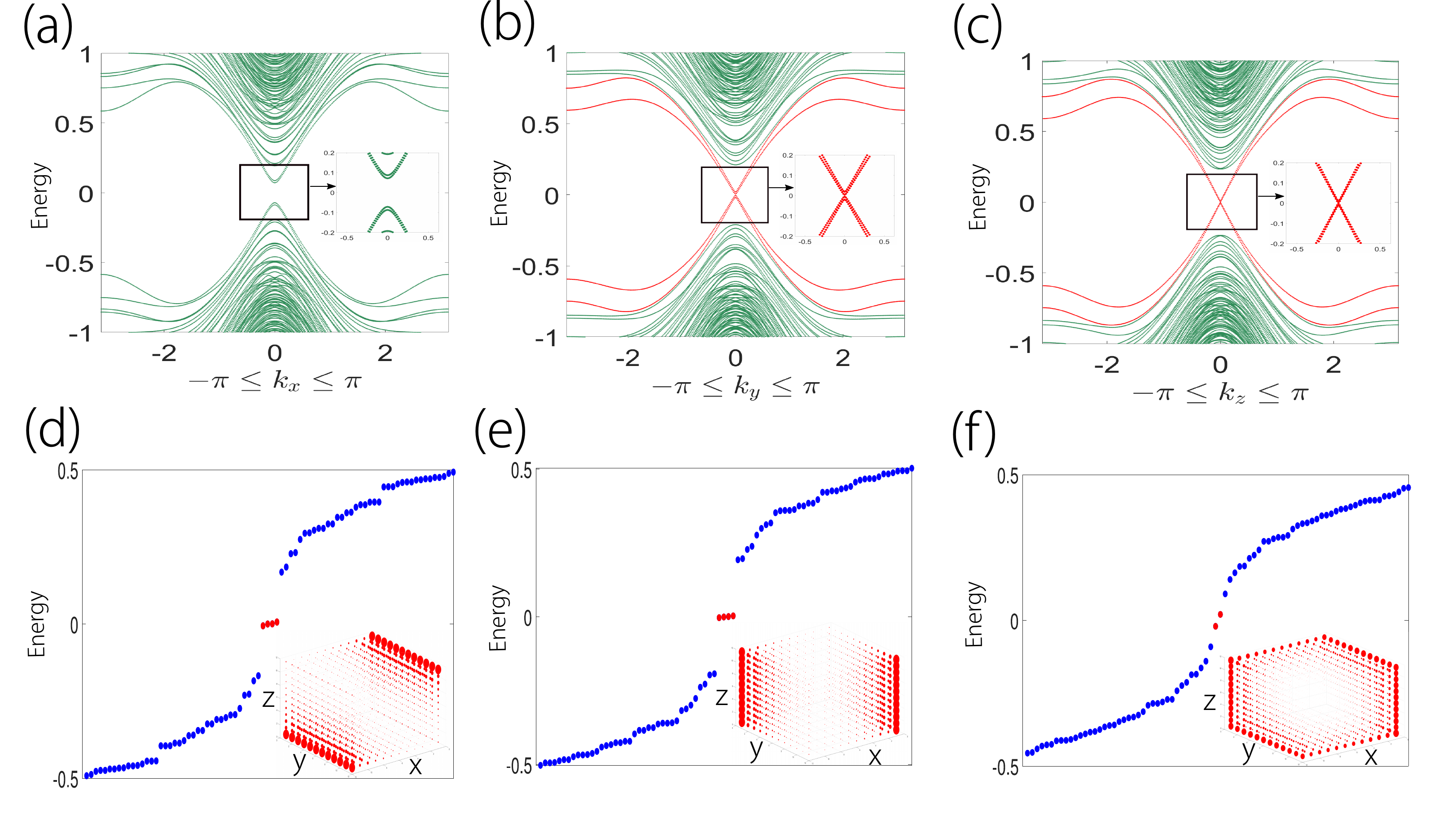}
	\caption{ The numerical results for Fig.~\ref{hierarchy}(f). The celluar diemnsion is $5$D$(4+1)$. The permitted perturbations are $i(\lambda_{14}\Gamma^1+\lambda_{24}\Gamma^2+\lambda_{34}\Gamma^3)\Gamma^4\Gamma^7+i(\lambda_{15}\Gamma^1+\lambda_{25}\Gamma^2+\lambda_{35}\Gamma^3)\Gamma^5\Gamma^7+\eta_1\Gamma^1\Gamma^6$, where $(\lambda_{14},\lambda_{15})=\alpha_1(\lambda_{24},\lambda_{25})=\alpha_2(\lambda_{24},\lambda_{25})$ and $\alpha_1,\alpha_2> 0$. Parameters are set as $\lambda_{14}=0.3,~\lambda_{24}=0.36,~\lambda_{34}=0.27,~\eta_1=0.1$. It is easy to verify that each surface is gapped. (a-c) show the energy spectra in the $k_{x/y/z}$ momentum space with the periodic boundary conditions along the $x/y/z$ axis, respectively. Hence, the edges along $y$ and $z$ directions could be gapless with helical modes. (d) and (e) show the energy spectrum in real space with the periodic boundary conditions for the $y$ and $z$ directions, respectively. The right inserts of (d) and (e) show the spatial distributions of in-gap states in real space with the periodic boundary conditions along $y$ and $z$ directions, respectively. The composition of THEM just gives Fig. \ref{hierarchy}(f).(f) shows the energy spectrum in real space with the open boundary conditions for all directions.  The right insert shows the spatial distribution of the states labeled by red dots.}
	\label{tenf}
\end{figure*}

\begin{figure*}[h]
	\includegraphics[width=7in]{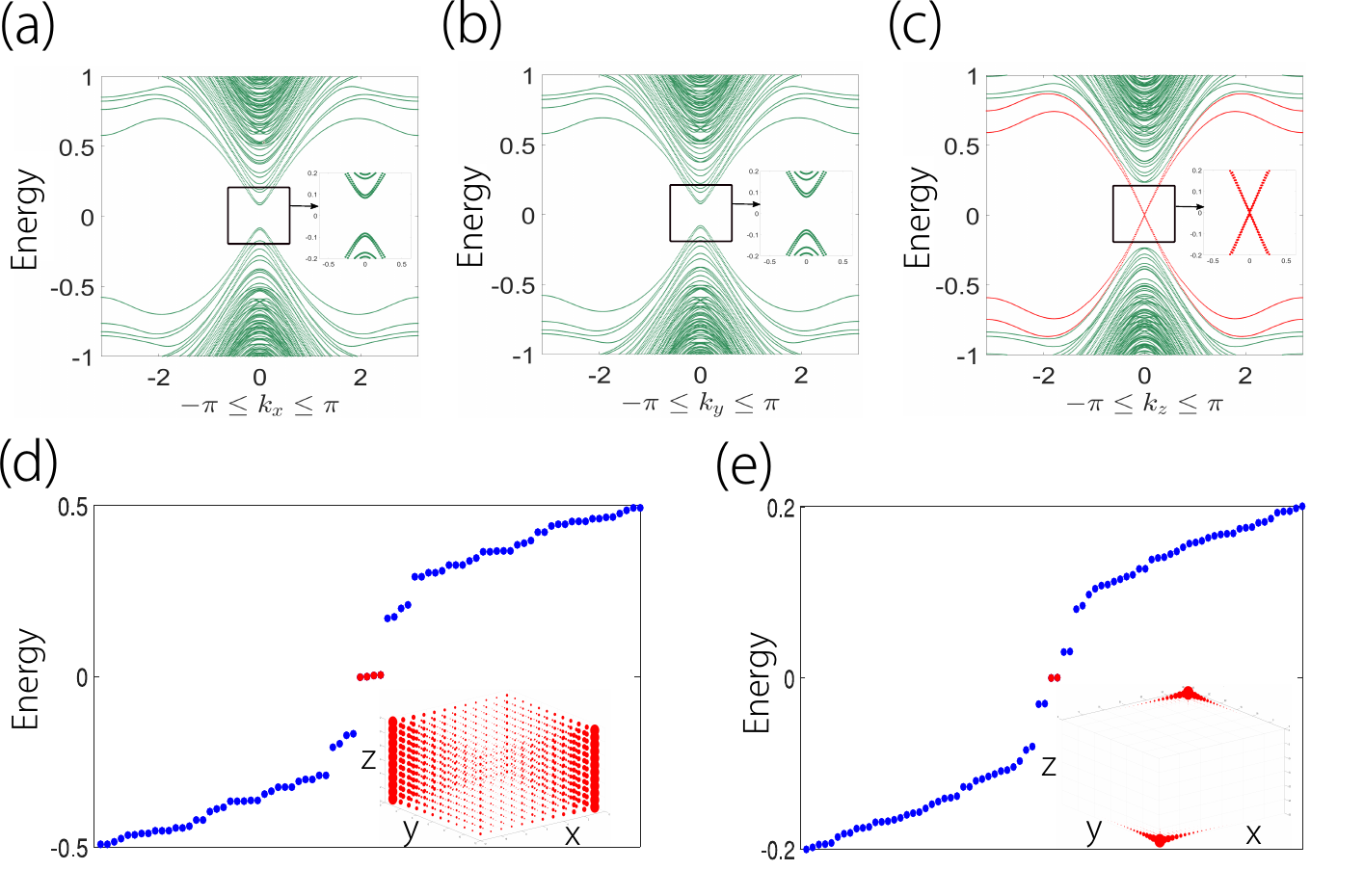}
	\caption{ The numerical results for Fig.~\ref{hierarchy}(g). The celluar dimension is $7$D$(5+2)$. The permitted perturbations are $i(\lambda_{14}\Gamma^1+\lambda_{24}\Gamma^2+\lambda_{34}\Gamma^3)\Gamma^4\Gamma^7+i(\lambda_{15}\Gamma^1+\lambda_{25}\Gamma^2+\lambda_{35}\Gamma^3)\Gamma^5\Gamma^7+(\eta_1\Gamma^1+\eta_2\Gamma^2)\Gamma^6$, where $(\lambda_{14},\lambda_{15})=\alpha(\lambda_{24},\lambda_{25}),~\alpha> 0,$ and $\eta_1\eta_2>0$. So the parameters are set as $\lambda_{14}=0.3,~\lambda_{24}=0.31,~\lambda_{34}=0.29,~\lambda_{15}=\lambda_{25}=0,~\lambda_{35}=0.01,~\eta_1=0.12,\eta_2=0.13$. It is easy to verify that each surface is gapped. (a)-(c) illustrate the energy spectra in the $k_{x/y/z}$ momentum space with the periodic boundary conditions along the $x/y/z$ axis, respectively. (d) and (e) show the energy spectrum in real space with periodic and open boundary conditions for the $z$ direction, respectively. Hence, only the edges along the $z$-directions could be gapless with helical modes. The right inserts of (d) and (e) illustrate the spatial distribution of zero-mode states in real space with periodic boundary conditions along $z$ direction and the open boundary conditions along all directions, respectively. The composition of (f) and (g) is just Fig. \ref{hierarchy}(g).}
	\label{teng}
\end{figure*}

\begin{figure*}[h]
	\includegraphics[width=7in]{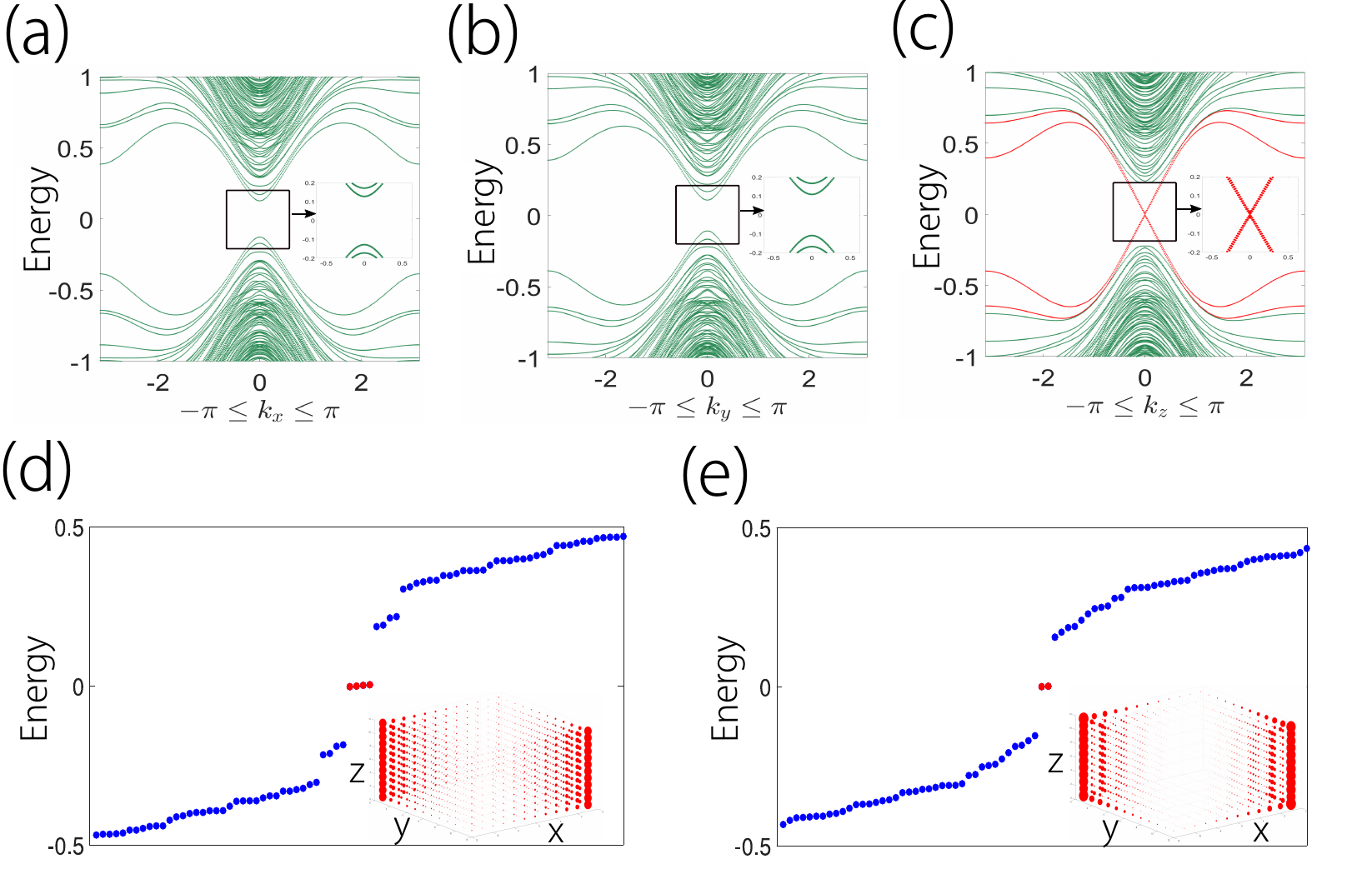}
	\caption{ The numerical results for Fig.~\ref{hierarchy}(h). The celluar diemnsion is $7$D$(5+2)$. The permitted perturbations are $i(\lambda_{14}\Gamma^1+\lambda_{24}\Gamma^2+\lambda_{34}\Gamma^3)\Gamma^4\Gamma^7+i(\lambda_{15}\Gamma^1+\lambda_{25}\Gamma^2+\lambda_{35}\Gamma^3)\Gamma^5\Gamma^7+(\eta_1\Gamma^1+\eta_2\Gamma^2)\Gamma^6$, where $(\lambda_{14},\lambda_{15})=\alpha(\lambda_{24},\lambda_{25}),~\alpha> 0,$ and $\eta_1\eta_2<0$. So the parameters are set as $\lambda_{14}=0.3,~\lambda_{24}=0.35,~\lambda_{34}=0.2,~\lambda_{15}=\lambda_{25}=0,~\lambda_{35}=0.21,~\eta_1=0.1,\eta_2=-0.1$. It is easy to verify that each surface is gapped. (a)-(c) illustrate the energy spectrum in the $k_{x/y/z}$ momentum space with periodic boundary conditions along the $x/y/z$ axis, respectively. Hence, only the $z$-edges could be gapless with helical modes. (d) and (e) show the energy spectrum in real space with the periodic boundary conditions for the $z$ direction and open boundary conditions for the other directions, respectively. The right inserts of (d) and (e) show the spatial distribution of in-gap states with the periodic boundary conditions along $z$ direction and open boundary conditions for all directions, respectively. Apparently, there are only a pair of edges parallel to the $z$ axis hosting helical modes.}
	\label{tenh}
\end{figure*}

\begin{figure*}[h]
	\includegraphics[width=7in]{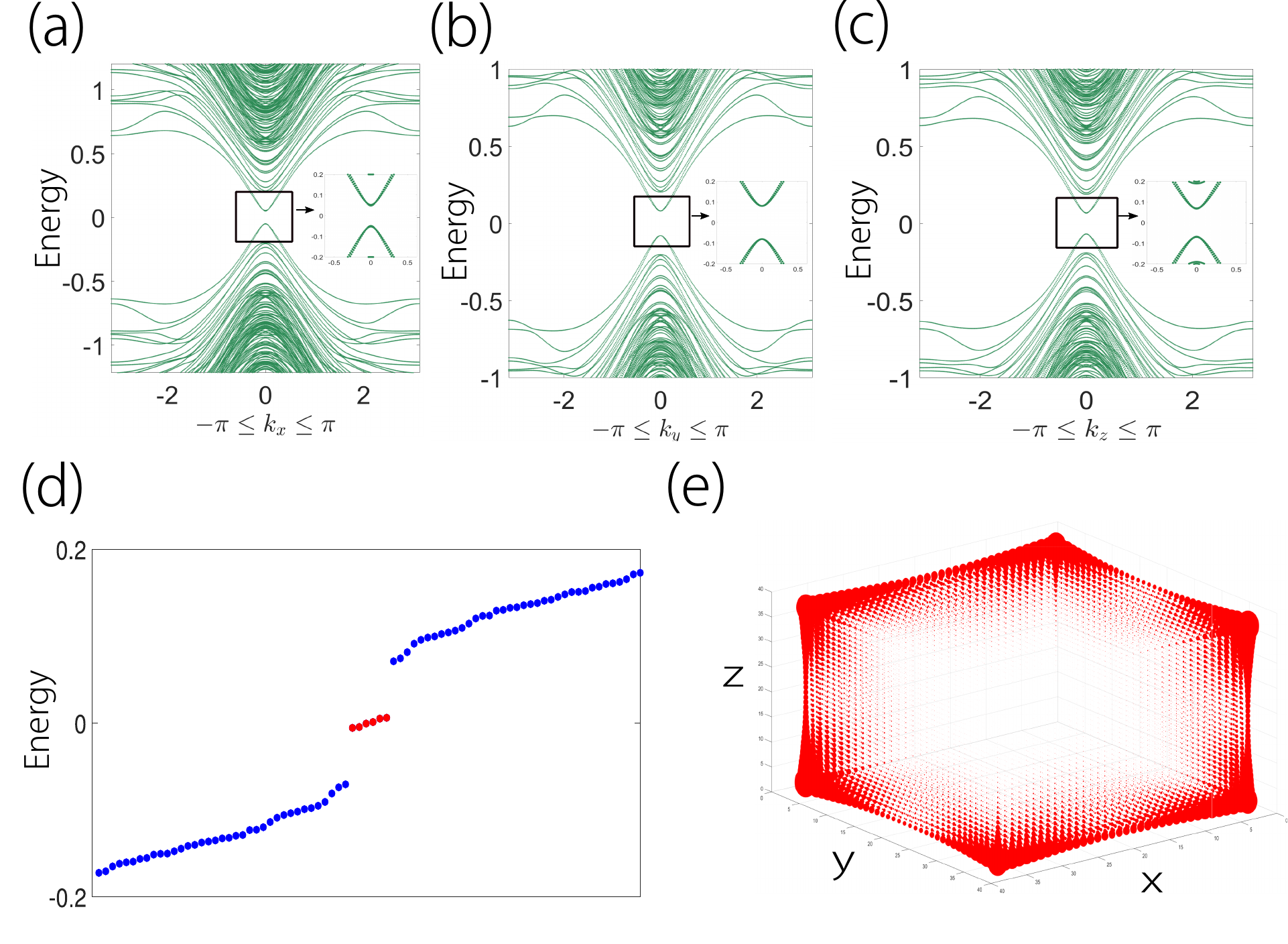}
	\caption{ The numerical results for Fig.~\ref{hierarchy}(i). The celluar diemnsion is $9$D$(6+3)$. The permitted perturbations are $i(\lambda_{14}\Gamma^1+\lambda_{24}\Gamma^2+\lambda_{34}\Gamma^3)\Gamma^4\Gamma^7+i(\lambda_{15}\Gamma^1+\lambda_{25}\Gamma^2+\lambda_{35}\Gamma^3)\Gamma^5\Gamma^7+(\eta_1\Gamma^1+\eta_2\Gamma^2+\eta_3\Gamma^3)\Gamma^6$. So the parameters are set as $\lambda_{14}=0.29,~\lambda_{24}=0.3,~\lambda_{34}=0.31,~\lambda_{15}=0.02,~\lambda_{25}=0.01,~\lambda_{35}=-0.01,~\eta_1=0.1,~\eta_2=0.11,~\eta_3=0.12$. It is easy to verify that each surface is gapped. (a)-(c) show the energy spectra in the $k_{x/y/z}$ momentum space with the periodic boundary conditions along the $x/y/z$ axis, respectively. (d) and (e) show the energy spectra and the spatial distribution of six in-gap states, respectively.}
	\label{teni}
\end{figure*}

\begin{figure*}[h]
	\includegraphics[width=7in]{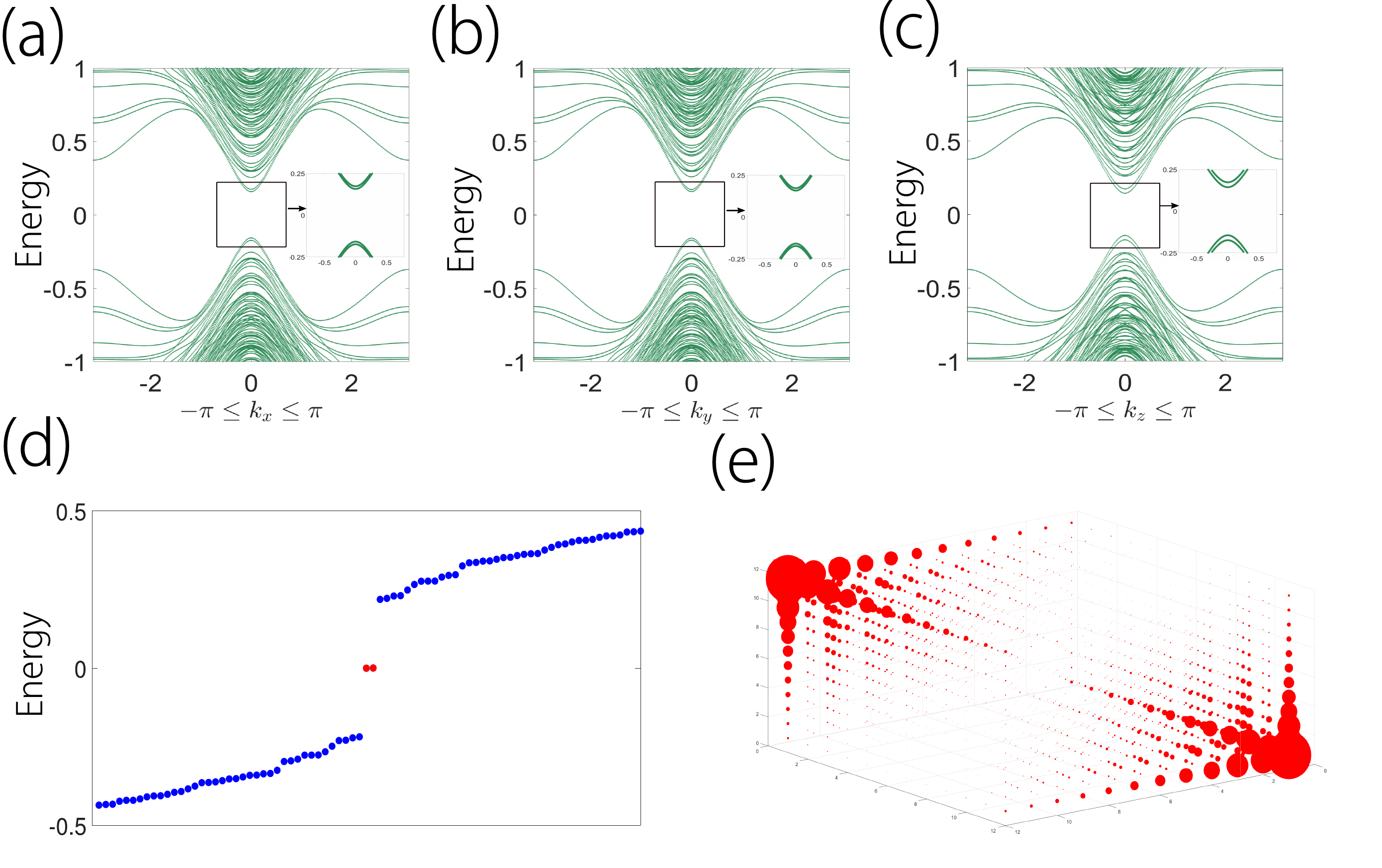}
	\caption{ The numerical results for Fig.~\ref{hierarchy}(j). The celluar diemnsion is $9$D$(6+3)$. The permitted perturbations are $i(\lambda_{14}\Gamma^1+\lambda_{24}\Gamma^2+\lambda_{34}\Gamma^3)\Gamma^4\Gamma^7+i(\lambda_{15}\Gamma^1+\lambda_{25}\Gamma^2+\lambda_{35}\Gamma^3)\Gamma^5\Gamma^7+(\eta_1\Gamma^1+\eta_2\Gamma^2+\eta_3\Gamma^3)\Gamma^6$. So the parameters are set as $\lambda_{14}=0.32,~\lambda_{24}=0.03,~\lambda_{34}=-0.26,~\lambda_{15}=0.04,~\lambda_{25}=0.3,~\lambda_{35}=0.25,~\eta_1=0.05,~\eta_2=-0.04,~\eta_3=0.02$. It is easy to verify that each surface is gapped. (a)-(c) illustrate the energy spectra in the $k_{x/y/z}$ momentum space with the periodic boundary conditions along the $x/y/z$ axis, respectively. Hence, all edges are also gapped. (d) and (e) show the energy spectra and the distributions of zero-energy states, respectively, in real space with the open boundary conditions for all directions.}
	\label{tenj}
\end{figure*}

\begin{figure*}[h]
	\includegraphics[width=7in]{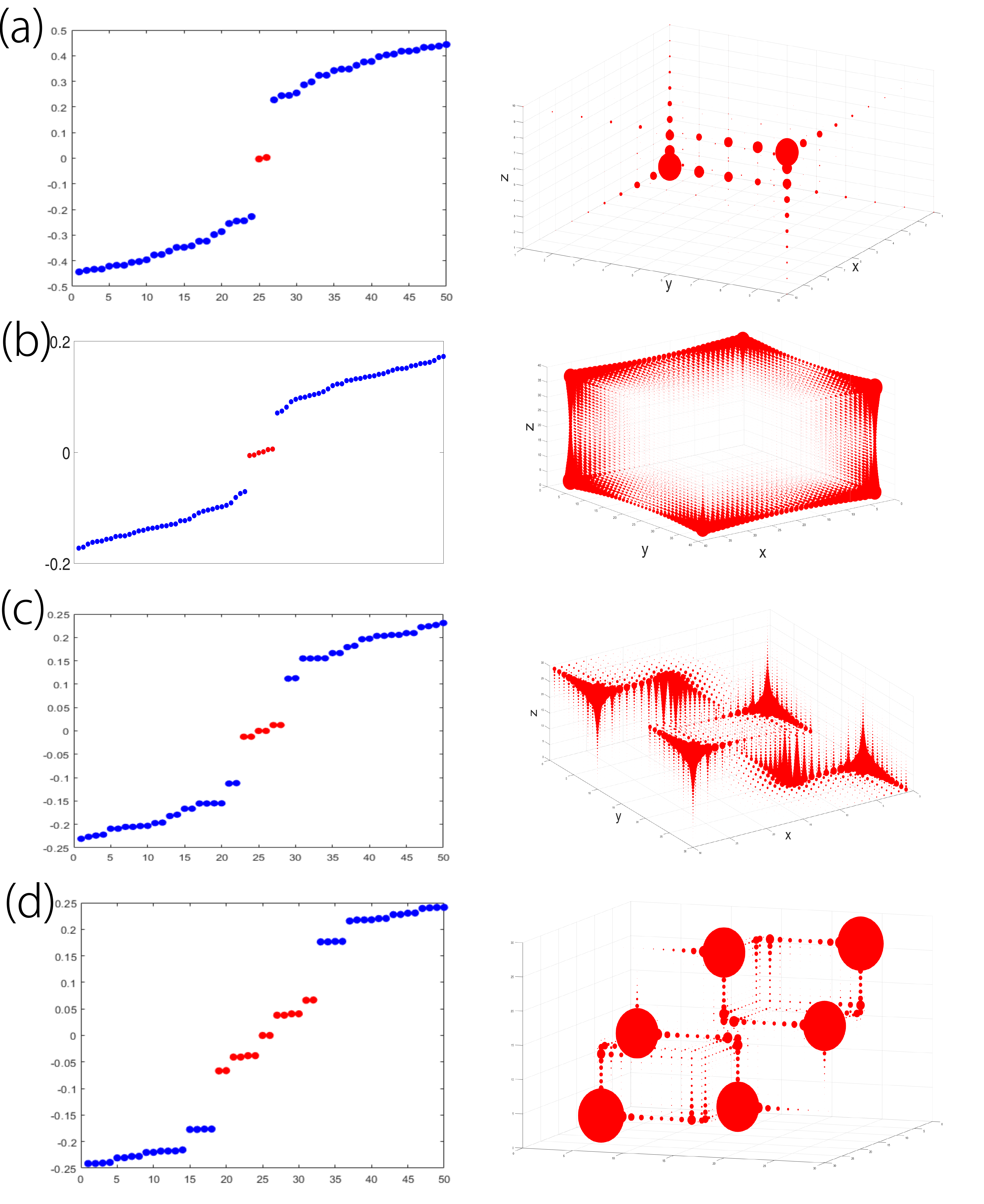}
	\caption{ The numerical results for FIG.~2 in the maintext. The parameters are set as $\lambda_{14}=0.3,~\lambda_{24}=-0.2,~\lambda_{34}=-0.2,~\lambda_{15}=0~\lambda_{25}=0.25,~\lambda_{35}=-0.25,~\eta_1=0,~\eta_2=0,~\eta_3=0$ in (a), (c) and (d). (b) corresponds to FIG.~\ref{teni}(d) and (e).}
	\label{simulation}
\end{figure*}
\clearpage
\newpage

\bibliographystyle{apsrev}
\bibliography{Takagi-Ref}

\begin{thebibliography}{41}
\expandafter\ifx\csname natexlab\endcsname\relax\def\natexlab#1{#1}\fi
\expandafter\ifx\csname bibnamefont\endcsname\relax
  \def\bibnamefont#1{#1}\fi
\expandafter\ifx\csname bibfnamefont\endcsname\relax
  \def\bibfnamefont#1{#1}\fi
\expandafter\ifx\csname citenamefont\endcsname\relax
  \def\citenamefont#1{#1}\fi
\expandafter\ifx\csname url\endcsname\relax
  \def\url#1{\texttt{#1}}\fi
\expandafter\ifx\csname urlprefix\endcsname\relax\def\urlprefix{URL }\fi
\providecommand{\bibinfo}[2]{#2}
\providecommand{\eprint}[2][]{\url{#2}}

\bibitem[{\citenamefont{Hasan and Kane}(2010)}]{Kane-RMP}
\bibinfo{author}{\bibfnamefont{M.~Z.} \bibnamefont{Hasan}} \bibnamefont{and}
  \bibinfo{author}{\bibfnamefont{C.~L.} \bibnamefont{Kane}},
  \bibinfo{journal}{Rev. Mod. Phys.} \textbf{\bibinfo{volume}{82}},
  \bibinfo{pages}{3045} (\bibinfo{year}{2010}).

\bibitem[{\citenamefont{Qi and Zhang}(2011)}]{XLQi-RMP}
\bibinfo{author}{\bibfnamefont{X.-L.} \bibnamefont{Qi}} \bibnamefont{and}
  \bibinfo{author}{\bibfnamefont{S.-C.} \bibnamefont{Zhang}},
  \bibinfo{journal}{Rev. Mod. Phys.} \textbf{\bibinfo{volume}{83}},
  \bibinfo{pages}{1057} (\bibinfo{year}{2011}).

\bibitem[{\citenamefont{Chiu et~al.}(2016)\citenamefont{Chiu, Teo, Schnyder,
  and Ryu}}]{ShinseiRyu-RMP}
\bibinfo{author}{\bibfnamefont{C.-K.} \bibnamefont{Chiu}},
  \bibinfo{author}{\bibfnamefont{J.~C.~Y.} \bibnamefont{Teo}},
  \bibinfo{author}{\bibfnamefont{A.~P.} \bibnamefont{Schnyder}},
  \bibnamefont{and} \bibinfo{author}{\bibfnamefont{S.}~\bibnamefont{Ryu}},
  \bibinfo{journal}{Rev. Mod. Phys.} \textbf{\bibinfo{volume}{88}},
  \bibinfo{pages}{035005} (\bibinfo{year}{2016}).

\bibitem[{\citenamefont{Volovik}(2003)}]{Volovik:book}
\bibinfo{author}{\bibfnamefont{G.~E.} \bibnamefont{Volovik}},
  \emph{\bibinfo{title}{Universe in a helium droplet}}
  (\bibinfo{publisher}{Oxford University Press, Oxford UK},
  \bibinfo{year}{2003}), ISBN \bibinfo{isbn}{0521670535}.

\bibitem[{\citenamefont{Shen}(2012)}]{Shun-Qing-Shen}
\bibinfo{author}{\bibfnamefont{S.-Q.} \bibnamefont{Shen}},
  \emph{\bibinfo{title}{Topological Insulators: Dirac Equation in Condensed
  Matters}} (\bibinfo{publisher}{Springer}, \bibinfo{year}{2012}).

\bibitem[{\citenamefont{Atiyah}(1966)}]{Atiyah-KR}
\bibinfo{author}{\bibfnamefont{M.~F.} \bibnamefont{Atiyah}},
  \bibinfo{journal}{The Quarterly Journal of Mathematics}
  \textbf{\bibinfo{volume}{17}}, \bibinfo{pages}{367} (\bibinfo{year}{1966}).

\bibitem[{\citenamefont{Schnyder et~al.}(2008)\citenamefont{Schnyder, Ryu,
  Furusaki, and Ludwig}}]{Schnyder2008}
\bibinfo{author}{\bibfnamefont{A.~P.} \bibnamefont{Schnyder}},
  \bibinfo{author}{\bibfnamefont{S.}~\bibnamefont{Ryu}},
  \bibinfo{author}{\bibfnamefont{A.}~\bibnamefont{Furusaki}}, \bibnamefont{and}
  \bibinfo{author}{\bibfnamefont{A.~W.~W.} \bibnamefont{Ludwig}},
  \bibinfo{journal}{Phys. Rev. B} \textbf{\bibinfo{volume}{78}},
  \bibinfo{pages}{195125} (\bibinfo{year}{2008}).

\bibitem[{\citenamefont{Kitaev}(2010)}]{Kitaev2009AIP}
\bibinfo{author}{\bibfnamefont{A.}~\bibnamefont{Kitaev}}, \bibinfo{journal}{AIP
  Conference Proceedings} \textbf{\bibinfo{volume}{1134}}, \bibinfo{pages}{22}
  (\bibinfo{year}{2010}).

\bibitem[{\citenamefont{Zhao and Wang}(2013)}]{ZhaoYXWang13prl}
\bibinfo{author}{\bibfnamefont{Y.~X.} \bibnamefont{Zhao}} \bibnamefont{and}
  \bibinfo{author}{\bibfnamefont{Z.~D.} \bibnamefont{Wang}},
  \bibinfo{journal}{Phys. Rev. Lett.} \textbf{\bibinfo{volume}{110}},
  \bibinfo{pages}{240404} (\bibinfo{year}{2013}).

\bibitem[{\citenamefont{Altland and Zirnbauer}(1997)}]{AZ-Classification}
\bibinfo{author}{\bibfnamefont{A.}~\bibnamefont{Altland}} \bibnamefont{and}
  \bibinfo{author}{\bibfnamefont{M.~R.} \bibnamefont{Zirnbauer}},
  \bibinfo{journal}{Phys. Rev. B} \textbf{\bibinfo{volume}{55}},
  \bibinfo{pages}{1142} (\bibinfo{year}{1997}).

\bibitem[{\citenamefont{Zhao and Wang}(2014)}]{ZhaoYXWang14Septprb}
\bibinfo{author}{\bibfnamefont{Y.~X.} \bibnamefont{Zhao}} \bibnamefont{and}
  \bibinfo{author}{\bibfnamefont{Z.~D.} \bibnamefont{Wang}},
  \bibinfo{journal}{Phys. Rev. B} \textbf{\bibinfo{volume}{89}},
  \bibinfo{pages}{075111} (\bibinfo{year}{2014}).

\bibitem[{\citenamefont{Zhang et~al.}(2013)\citenamefont{Zhang, Kane, and
  Mele}}]{ZhangPRL2013}
\bibinfo{author}{\bibfnamefont{F.}~\bibnamefont{Zhang}},
  \bibinfo{author}{\bibfnamefont{C.~L.} \bibnamefont{Kane}}, \bibnamefont{and}
  \bibinfo{author}{\bibfnamefont{E.~J.} \bibnamefont{Mele}},
  \bibinfo{journal}{Phys. Rev. Lett.} \textbf{\bibinfo{volume}{110}},
  \bibinfo{pages}{046404} (\bibinfo{year}{2013}).

\bibitem[{\citenamefont{Benalcazar et~al.}(2017)\citenamefont{Benalcazar,
  Bernevig, and Hughes}}]{Benalcazar61}
\bibinfo{author}{\bibfnamefont{W.~A.} \bibnamefont{Benalcazar}},
  \bibinfo{author}{\bibfnamefont{B.~A.} \bibnamefont{Bernevig}},
  \bibnamefont{and} \bibinfo{author}{\bibfnamefont{T.~L.}
  \bibnamefont{Hughes}}, \bibinfo{journal}{Science}
  \textbf{\bibinfo{volume}{357}}, \bibinfo{pages}{61} (\bibinfo{year}{2017}),
  ISSN \bibinfo{issn}{0036-8075}.

\bibitem[{\citenamefont{Langbehn et~al.}(2017)\citenamefont{Langbehn, Peng,
  Trifunovic, von Oppen, and Brouwer}}]{LangPRL2017}
\bibinfo{author}{\bibfnamefont{J.}~\bibnamefont{Langbehn}},
  \bibinfo{author}{\bibfnamefont{Y.}~\bibnamefont{Peng}},
  \bibinfo{author}{\bibfnamefont{L.}~\bibnamefont{Trifunovic}},
  \bibinfo{author}{\bibfnamefont{F.}~\bibnamefont{von Oppen}},
  \bibnamefont{and} \bibinfo{author}{\bibfnamefont{P.~W.}
  \bibnamefont{Brouwer}}, \bibinfo{journal}{Phys. Rev. Lett.}
  \textbf{\bibinfo{volume}{119}}, \bibinfo{pages}{246401}
  (\bibinfo{year}{2017}).

\bibitem[{\citenamefont{Song et~al.}(2017)\citenamefont{Song, Fang, and
  Fang}}]{SongPRL2017}
\bibinfo{author}{\bibfnamefont{Z.}~\bibnamefont{Song}},
  \bibinfo{author}{\bibfnamefont{Z.}~\bibnamefont{Fang}}, \bibnamefont{and}
  \bibinfo{author}{\bibfnamefont{C.}~\bibnamefont{Fang}},
  \bibinfo{journal}{Phys. Rev. Lett.} \textbf{\bibinfo{volume}{119}},
  \bibinfo{pages}{246402} (\bibinfo{year}{2017}).

\bibitem[{\citenamefont{Zhao et~al.}(2016)\citenamefont{Zhao, Schnyder, and
  Wang}}]{ZhaoWang16Aprprl}
\bibinfo{author}{\bibfnamefont{Y.~X.} \bibnamefont{Zhao}},
  \bibinfo{author}{\bibfnamefont{A.~P.} \bibnamefont{Schnyder}},
  \bibnamefont{and} \bibinfo{author}{\bibfnamefont{Z.~D.} \bibnamefont{Wang}},
  \bibinfo{journal}{Phys. Rev. Lett.} \textbf{\bibinfo{volume}{116}},
  \bibinfo{pages}{156402} (\bibinfo{year}{2016}).

\bibitem[{\citenamefont{Kruthoff et~al.}(2017)\citenamefont{Kruthoff, de~Boer,
  van Wezel, Kane, and Slager}}]{Band-Combinatorics}
\bibinfo{author}{\bibfnamefont{J.}~\bibnamefont{Kruthoff}},
  \bibinfo{author}{\bibfnamefont{J.}~\bibnamefont{de~Boer}},
  \bibinfo{author}{\bibfnamefont{J.}~\bibnamefont{van Wezel}},
  \bibinfo{author}{\bibfnamefont{C.~L.} \bibnamefont{Kane}}, \bibnamefont{and}
  \bibinfo{author}{\bibfnamefont{R.-J.} \bibnamefont{Slager}},
  \bibinfo{journal}{Phys. Rev. X} \textbf{\bibinfo{volume}{7}},
  \bibinfo{pages}{041069} (\bibinfo{year}{2017}).

\bibitem[{\citenamefont{Zhao and Lu}(2017)}]{ZhaoLu17Aprprl}
\bibinfo{author}{\bibfnamefont{Y.~X.} \bibnamefont{Zhao}} \bibnamefont{and}
  \bibinfo{author}{\bibfnamefont{Y.}~\bibnamefont{Lu}}, \bibinfo{journal}{Phys.
  Rev. Lett.} \textbf{\bibinfo{volume}{118}}, \bibinfo{pages}{056401}
  (\bibinfo{year}{2017}).

\bibitem[{\citenamefont{Kim et~al.}(2015)\citenamefont{Kim, Wieder, Kane, and
  Rappe}}]{PhysRevLett.115.036806}
\bibinfo{author}{\bibfnamefont{Y.}~\bibnamefont{Kim}},
  \bibinfo{author}{\bibfnamefont{B.~J.} \bibnamefont{Wieder}},
  \bibinfo{author}{\bibfnamefont{C.~L.} \bibnamefont{Kane}}, \bibnamefont{and}
  \bibinfo{author}{\bibfnamefont{A.~M.} \bibnamefont{Rappe}},
  \bibinfo{journal}{Phys. Rev. Lett.} \textbf{\bibinfo{volume}{115}},
  \bibinfo{pages}{036806} (\bibinfo{year}{2015}).

\bibitem[{\citenamefont{Yu et~al.}(2015)\citenamefont{Yu, Weng, Fang, Dai, and
  Hu}}]{PhysRevLett.115.036807}
\bibinfo{author}{\bibfnamefont{R.}~\bibnamefont{Yu}},
  \bibinfo{author}{\bibfnamefont{H.}~\bibnamefont{Weng}},
  \bibinfo{author}{\bibfnamefont{Z.}~\bibnamefont{Fang}},
  \bibinfo{author}{\bibfnamefont{X.}~\bibnamefont{Dai}}, \bibnamefont{and}
  \bibinfo{author}{\bibfnamefont{X.}~\bibnamefont{Hu}}, \bibinfo{journal}{Phys.
  Rev. Lett.} \textbf{\bibinfo{volume}{115}}, \bibinfo{pages}{036807}
  (\bibinfo{year}{2015}).

\bibitem[{\citenamefont{Chan et~al.}(2016)\citenamefont{Chan, Chiu, Chou, and
  Schnyder}}]{PhysRevB.93.205132}
\bibinfo{author}{\bibfnamefont{Y.-H.} \bibnamefont{Chan}},
  \bibinfo{author}{\bibfnamefont{C.-K.} \bibnamefont{Chiu}},
  \bibinfo{author}{\bibfnamefont{M.~Y.} \bibnamefont{Chou}}, \bibnamefont{and}
  \bibinfo{author}{\bibfnamefont{A.~P.} \bibnamefont{Schnyder}},
  \bibinfo{journal}{Phys. Rev. B} \textbf{\bibinfo{volume}{93}},
  \bibinfo{pages}{205132} (\bibinfo{year}{2016}).

\bibitem[{\citenamefont{Ahn et~al.}(2019)\citenamefont{Ahn, Park, and
  Yang}}]{B-J-Yang19APRPRX}
\bibinfo{author}{\bibfnamefont{J.}~\bibnamefont{Ahn}},
  \bibinfo{author}{\bibfnamefont{S.}~\bibnamefont{Park}}, \bibnamefont{and}
  \bibinfo{author}{\bibfnamefont{B.-J.} \bibnamefont{Yang}},
  \bibinfo{journal}{Phys. Rev. X} \textbf{\bibinfo{volume}{9}},
  \bibinfo{pages}{021013} (\bibinfo{year}{2019}).

\bibitem[{\citenamefont{Sheng et~al.}(2019)\citenamefont{Sheng, Chen, Liu,
  Chen, Yu, Zhao, and Yang}}]{ZhaoYang19prl}
\bibinfo{author}{\bibfnamefont{X.-L.} \bibnamefont{Sheng}},
  \bibinfo{author}{\bibfnamefont{C.}~\bibnamefont{Chen}},
  \bibinfo{author}{\bibfnamefont{H.}~\bibnamefont{Liu}},
  \bibinfo{author}{\bibfnamefont{Z.}~\bibnamefont{Chen}},
  \bibinfo{author}{\bibfnamefont{Z.-M.} \bibnamefont{Yu}},
  \bibinfo{author}{\bibfnamefont{Y.~X.} \bibnamefont{Zhao}}, \bibnamefont{and}
  \bibinfo{author}{\bibfnamefont{S.~A.} \bibnamefont{Yang}},
  \bibinfo{journal}{Phys. Rev. Lett.} \textbf{\bibinfo{volume}{123}},
  \bibinfo{pages}{256402} (\bibinfo{year}{2019}).

\bibitem[{\citenamefont{Wu et~al.}(2019)\citenamefont{Wu, Soluyanov, and
  Bzdu{\v s}ek}}]{Wu1273}
\bibinfo{author}{\bibfnamefont{Q.}~\bibnamefont{Wu}},
  \bibinfo{author}{\bibfnamefont{A.~A.} \bibnamefont{Soluyanov}},
  \bibnamefont{and} \bibinfo{author}{\bibfnamefont{T.}~\bibnamefont{Bzdu{\v
  s}ek}}, \bibinfo{journal}{Science} \textbf{\bibinfo{volume}{365}},
  \bibinfo{pages}{1273} (\bibinfo{year}{2019}).

\bibitem[{\citenamefont{Wang et~al.}(2019)\citenamefont{Wang, Wieder, Li, Yan,
  and Bernevig}}]{Wangzhijun2019prl}
\bibinfo{author}{\bibfnamefont{Z.}~\bibnamefont{Wang}},
  \bibinfo{author}{\bibfnamefont{B.~J.} \bibnamefont{Wieder}},
  \bibinfo{author}{\bibfnamefont{J.}~\bibnamefont{Li}},
  \bibinfo{author}{\bibfnamefont{B.}~\bibnamefont{Yan}}, \bibnamefont{and}
  \bibinfo{author}{\bibfnamefont{B.~A.} \bibnamefont{Bernevig}},
  \bibinfo{journal}{Phys. Rev. Lett.} \textbf{\bibinfo{volume}{123}},
  \bibinfo{pages}{186401} (\bibinfo{year}{2019}).

\bibitem[{\citenamefont{Li et~al.}(2020)\citenamefont{Li, Mekawy, Krasnok, and
  Al\`u}}]{PhysRevLett.124.193901}
\bibinfo{author}{\bibfnamefont{H.}~\bibnamefont{Li}},
  \bibinfo{author}{\bibfnamefont{A.}~\bibnamefont{Mekawy}},
  \bibinfo{author}{\bibfnamefont{A.}~\bibnamefont{Krasnok}}, \bibnamefont{and}
  \bibinfo{author}{\bibfnamefont{A.}~\bibnamefont{Al\`u}},
  \bibinfo{journal}{Phys. Rev. Lett.} \textbf{\bibinfo{volume}{124}},
  \bibinfo{pages}{193901} (\bibinfo{year}{2020}).

\bibitem[{\citenamefont{Ahn et~al.}(2018)\citenamefont{Ahn, Kim, Kim, and
  Yang}}]{AhnPRL2018}
\bibinfo{author}{\bibfnamefont{J.}~\bibnamefont{Ahn}},
  \bibinfo{author}{\bibfnamefont{D.}~\bibnamefont{Kim}},
  \bibinfo{author}{\bibfnamefont{Y.}~\bibnamefont{Kim}}, \bibnamefont{and}
  \bibinfo{author}{\bibfnamefont{B.-J.} \bibnamefont{Yang}},
  \bibinfo{journal}{Phys. Rev. Lett.} \textbf{\bibinfo{volume}{121}},
  \bibinfo{pages}{106403} (\bibinfo{year}{2018}).

\bibitem[{\citenamefont{Takagi}(1924)}]{Takagi}
\bibinfo{author}{\bibfnamefont{T.}~\bibnamefont{Takagi}},
  \bibinfo{journal}{Japanese J. Math} \textbf{\bibinfo{volume}{1}},
  \bibinfo{pages}{83} (\bibinfo{year}{1924}).

\bibitem[{Tak()}]{Takaji_wiki}
\bibinfo{note}{See sec 3.6 of the wikipedia item:
  \url{https://en.wikipedia.org/wiki/Matrix_decomposition}}.

\bibitem[{PT-()}]{PT-symmetry}
\bibinfo{note}{Once $(\hat{P}\hat{T})^2=1$ is satisfied, there always exists a
  unitary transformation $U$, such that
  $UU_{PT}\hat{\mathcal{K}}U^\dagger=e^{i\phi}\sigma_1\hat{\mathcal{K}}$. But
  $e^{i\phi}\sigma_1\hat{\mathcal{K}}$ is equivalent to
  $\sigma_1\hat{\mathcal{K}}$, since they impose the same constraint to the
  Hamiltonian.}

\bibitem[{\citenamefont{Bernevig and Hughes}(2013)}]{bernevig2013topological}
\bibinfo{author}{\bibfnamefont{B.}~\bibnamefont{Bernevig}} \bibnamefont{and}
  \bibinfo{author}{\bibfnamefont{T.}~\bibnamefont{Hughes}},
  \emph{\bibinfo{title}{Topological Insulators and Topological
  Superconductors}} (\bibinfo{publisher}{Princeton University Press},
  \bibinfo{year}{2013}), ISBN \bibinfo{isbn}{9780691151755}.

\bibitem[{Glo()}]{Global-Torus_Lifting}
\bibinfo{note}{Strictly speaking, we assume no topologically nontrivial
  configuration over lower-dimensional tori of the Brillouin torus $T^3$.}

\bibitem[{Com()}]{Comparison_AIII}
\bibinfo{note}{It is noteworthy the essential distinction of the topological
  invariant from that for the 3D TI in class AIII~\cite{Schnyder2008}, despite
  of the appearance similarity. The latter concerns $\mathcal{Q}(\k)$, while
  the present case concerns the Takagi factor $\U(\k)$ of $\mathcal{Q}(\k)$.}

\bibitem[{Sti()}]{Stiefel-Whitney}
\bibinfo{note}{The second Stiefel-Whitney number, named as the real Chern
  number in Ref.~\cite{ZhaoLu17Aprprl}, is a $\Z_2$ invariant applying for
  $\P\T$-invariant real systems, for which the transition function of real
  valence wavefunctions over $D_{N,S}^2$ is valued in $O(N)$. Hence, the
  fundamental group $\pi_1[O(N)]$ can impose obstructions for a globally
  defined basis of valence wavefunctions.}

\bibitem[{\citenamefont{Lu et~al.}(2014)\citenamefont{Lu, Joannopoulos, and
  Soljačić}}]{Lu2014}
\bibinfo{author}{\bibfnamefont{L.}~\bibnamefont{Lu}},
  \bibinfo{author}{\bibfnamefont{J.~D.} \bibnamefont{Joannopoulos}},
  \bibnamefont{and}
  \bibinfo{author}{\bibfnamefont{M.}~\bibnamefont{Soljačić}},
  \bibinfo{journal}{Nature Photonics} \textbf{\bibinfo{volume}{8}},
  \bibinfo{pages}{821} (\bibinfo{year}{2014}).

\bibitem[{\citenamefont{Yang et~al.}(2015)\citenamefont{Yang, Gao, Shi, Lin,
  Gao, Chong, and Zhang}}]{YangZjprl2015}
\bibinfo{author}{\bibfnamefont{Z.}~\bibnamefont{Yang}},
  \bibinfo{author}{\bibfnamefont{F.}~\bibnamefont{Gao}},
  \bibinfo{author}{\bibfnamefont{X.}~\bibnamefont{Shi}},
  \bibinfo{author}{\bibfnamefont{X.}~\bibnamefont{Lin}},
  \bibinfo{author}{\bibfnamefont{Z.}~\bibnamefont{Gao}},
  \bibinfo{author}{\bibfnamefont{Y.}~\bibnamefont{Chong}}, \bibnamefont{and}
  \bibinfo{author}{\bibfnamefont{B.}~\bibnamefont{Zhang}},
  \bibinfo{journal}{Phys. Rev. Lett.} \textbf{\bibinfo{volume}{114}},
  \bibinfo{pages}{114301} (\bibinfo{year}{2015}).

\bibitem[{\citenamefont{Mittal et~al.}(2019)\citenamefont{Mittal, Orre, Zhu,
  Gorlach, Poddubny, and Hafezi}}]{Mittal2019}
\bibinfo{author}{\bibfnamefont{S.}~\bibnamefont{Mittal}},
  \bibinfo{author}{\bibfnamefont{V.~V.} \bibnamefont{Orre}},
  \bibinfo{author}{\bibfnamefont{G.}~\bibnamefont{Zhu}},
  \bibinfo{author}{\bibfnamefont{M.~A.} \bibnamefont{Gorlach}},
  \bibinfo{author}{\bibfnamefont{A.}~\bibnamefont{Poddubny}}, \bibnamefont{and}
  \bibinfo{author}{\bibfnamefont{M.}~\bibnamefont{Hafezi}},
  \bibinfo{journal}{Nature Photonics} \textbf{\bibinfo{volume}{13}},
  \bibinfo{pages}{692} (\bibinfo{year}{2019}).

\bibitem[{\citenamefont{Xue et~al.}(2020)\citenamefont{Xue, Ge, Sun, Wang, Jia,
  Guan, Yuan, Chong, and Zhang}}]{Xue2020}
\bibinfo{author}{\bibfnamefont{H.}~\bibnamefont{Xue}},
  \bibinfo{author}{\bibfnamefont{Y.}~\bibnamefont{Ge}},
  \bibinfo{author}{\bibfnamefont{H.-X.} \bibnamefont{Sun}},
  \bibinfo{author}{\bibfnamefont{Q.}~\bibnamefont{Wang}},
  \bibinfo{author}{\bibfnamefont{D.}~\bibnamefont{Jia}},
  \bibinfo{author}{\bibfnamefont{Y.-J.} \bibnamefont{Guan}},
  \bibinfo{author}{\bibfnamefont{S.-Q.} \bibnamefont{Yuan}},
  \bibinfo{author}{\bibfnamefont{Y.}~\bibnamefont{Chong}}, \bibnamefont{and}
  \bibinfo{author}{\bibfnamefont{B.}~\bibnamefont{Zhang}},
  \bibinfo{journal}{Nature Communications} \textbf{\bibinfo{volume}{11}},
  \bibinfo{pages}{2442} (\bibinfo{year}{2020}).

\bibitem[{\citenamefont{Imhof et~al.}(2018)\citenamefont{Imhof, Berger, Bayer,
  Brehm, Molenkamp, Kiessling, Schindler, Lee, Greiter, Neupert
  et~al.}}]{Ronny_2018np}
\bibinfo{author}{\bibfnamefont{S.}~\bibnamefont{Imhof}},
  \bibinfo{author}{\bibfnamefont{C.}~\bibnamefont{Berger}},
  \bibinfo{author}{\bibfnamefont{F.}~\bibnamefont{Bayer}},
  \bibinfo{author}{\bibfnamefont{J.}~\bibnamefont{Brehm}},
  \bibinfo{author}{\bibfnamefont{L.~W.} \bibnamefont{Molenkamp}},
  \bibinfo{author}{\bibfnamefont{T.}~\bibnamefont{Kiessling}},
  \bibinfo{author}{\bibfnamefont{F.}~\bibnamefont{Schindler}},
  \bibinfo{author}{\bibfnamefont{C.~H.} \bibnamefont{Lee}},
  \bibinfo{author}{\bibfnamefont{M.}~\bibnamefont{Greiter}},
  \bibinfo{author}{\bibfnamefont{T.}~\bibnamefont{Neupert}},
  \bibnamefont{et~al.}, \bibinfo{journal}{Nature Physics}
  \textbf{\bibinfo{volume}{14}}, \bibinfo{pages}{925} (\bibinfo{year}{2018}).

\bibitem[{\citenamefont{Yu et~al.}(2020)\citenamefont{Yu, Zhao, and
  Schnyder}}]{Yu_Zhao_NSR}
\bibinfo{author}{\bibfnamefont{R.}~\bibnamefont{Yu}},
  \bibinfo{author}{\bibfnamefont{Y.~X.} \bibnamefont{Zhao}}, \bibnamefont{and}
  \bibinfo{author}{\bibfnamefont{A.~P.} \bibnamefont{Schnyder}},
  \bibinfo{journal}{National Science Review} \textbf{\bibinfo{volume}{7}},
  \bibinfo{pages}{1288} (\bibinfo{year}{2020}).

\bibitem[{\citenamefont{Huber}(2016)}]{Huber2020np}
\bibinfo{author}{\bibfnamefont{S.~D.} \bibnamefont{Huber}},
  \bibinfo{journal}{Nature Physics} \textbf{\bibinfo{volume}{12}},
  \bibinfo{pages}{621} (\bibinfo{year}{2016}).

\end{thebibliography}

\end{document}